\documentclass[aps,twocolumn,12pt,showpacs,showkeys,tightenlines,nofootinbib]{revtex4}

\usepackage{graphicx}
\usepackage{epsfig}\parskip 5pt plus 1pt
\usepackage{multirow}
\usepackage{subfigure}
\usepackage{enumerate}
\usepackage{amsmath}
\usepackage{amssymb}
\usepackage{amsfonts}
\usepackage{mathrsfs}
\usepackage{enumerate}
\usepackage{appendix}

\usepackage{amsbsy}
\usepackage{rotating}
\usepackage{color}

\definecolor{MyGrey}{rgb}{0,0,0} 
\definecolor{MyDarkBlue}{rgb}{0.23,0.21,0.69} 
\definecolor{MyLightBlue}{rgb}{0.22,0.51,0.86}

\usepackage[colorlinks=true,linkcolor=MyGrey,citecolor=MyGrey,urlcolor=MyGrey,bookmarksnumbered=true,bookmarksopen]{hyperref}

\newcommand{\be}{\begin{equation}}
\newcommand{\ee}{\end{equation}}
\newcommand{\bea}{\begin{eqnarray}}
\newcommand{\eea}{\end{eqnarray}}

\begin{document}

\begin{titlepage}

\renewcommand{\thefootnote}{\alph{footnote}}
\title{Neutrino-nucleus interaction models and their impact on oscillation analyses}

\author{P.~Coloma, P.~Huber, C.-M.~Jen and C.~Mariani
}
\affiliation{Center for Neutrino Physics, Virginia Tech, Blacksburg, VA 24061, USA}
\date{\today}
\pacs{14.60.Pq, 14.60.Lm}
\keywords{neutrino oscillation, neutrino cross section, final state interactions, nuclear effects}

\begin{abstract}
In neutrino oscillation experiments, neutrino interactions at the
detector are simulated using event generators which attempt to reflect
our understanding of nuclear physics. We study the impact of different
neutrino interactions and nuclear models on the determination of
neutrino oscillation parameters. We use two independent neutrino
event generators, GENIE and GiBUU, and apply them to a setup with
a conventional neutrino beam aiming at a water \v{C}erenkov detector,
for which only the QE-like sample is selected. Subsequently, we
perform a fit to the oscillation parameters in the $\nu_\mu$
disappearance channel.
\end{abstract}

\maketitle

\end{titlepage}

\newpage 

\section{Introduction}
\label{sec:intro}

Neutrino physics in the past two decades has seen an astounding
transformation from a collection of anomalies to precision science,
which most recently resulted in the measurement of
$\theta_{13}$ at accelerator~\cite{Abe:2011sj} and reactor~\cite{Abe:2011fz,An:2012eh,Ahn:2012nd} experiments. 
The next goals in neutrino oscillation physics are the determination of
the mass hierarchy (\textit{i.e.}, the ordering of the neutrino mass
eigenstates) and of the leptonic CP phase, which will require
control of systematic errors at an unprecedented level of accuracy in 
neutrino physics. In particular the measurement of the CP phase will 
put very stringent demands on the determination of neutrino versus 
antineutrino interactions in the GeV region, see {\it e.g.}
Refs.~\cite{Itow:2002rk,Harris:2004iq,Huber:2007em,FernandezMartinez:2010dm,Meloni:2012fq,Coloma:2012ji}. It
is widely recognized that our current understanding of nuclear effects
in neutrino-nucleus interactions is insufficient to guarantee the
required control of systematical errors. A number of studies have
been performed focusing on nuclear effects in this context, see for instance 
Refs.~\cite{Martini:2012fa,Nieves:2012yz,Lalakulich:2012hs, Martini:2012uc, Mosel:2013fxa}. 
Given the complexity of the full problem with respect to
neutrino and antineutrino cross-sections, we investigate here the
simpler case of $\nu_\mu$ disappearance. In particular, we would like
to extend the results of Ref.~\cite{Coloma:2013rqa} and perform a
comparison of \emph{different} nuclear models and their impact on
oscillation parameters. It is a common approach to estimate errors
arising from theories using several available theoretical calculations
and take the spread in their results as a measure of the associated
uncertainty; we follow the same logic and specifically, will compare
the results obtained using GENIE~\cite{Andreopoulos:2009rq} and
GiBUU~\cite{Leitner:2009ke,Buss:2011mx}. Clearly, this type of
inter-comparison can be extended to a larger number of event generators
as well. Our choice is largely guided by the fact that GENIE is widely
used, and GiBUU represents an unique and complementary theoretical
approach based on transport theory. Moreover, both codes are open
sources which seem a necessary condition for meaningful comparisons.
We evaluate the impact of several aspects of nuclear models and
their differing implementation with respect to the ability to measure
the so-called atmospheric parameters, $\Delta m^2_{31}$ and
$\theta_{23}$ in an experimental setup which is similar to T2K. Our
main result is that effects from changing the target nucleus from carbon to
oxygen induces a bias in $\Delta m^2_{31}$ of about
1\,$\sigma$. Fitting data obtained with one generator and the other
results in dramatic shifts due to different modeling and implementation of
final state interactions and nuclear models. Also, the absence of multi-nucleon
correlations in the fit can induce a bias between 1\,$\sigma$ and
3\,$\sigma$ on the results for both of the oscillation parameters. As 
previously found in Ref.~\cite{Coloma:2013rqa}, a near detector does 
not resolve these issues.
\par The paper is organized as follows. In Sec.~\ref{sec:events}, we
outline the principle of energy reconstruction for charged-current 
quasi-elastic-like events, and in Sec.~\ref{sec:gen} we perform a
detailed comparison of the various physics models implemented
in both generators for a number of relevant interaction models. This 
is followed by a detailed description of the simulations in
Sec.~\ref{sec:sim} which leads to our results in
Sec.~\ref{sec:results}. Finally, in Sec.~\ref{sec:concl} we
present our conclusions.
%
\section{The QE-like event sample}
\label{sec:events}

In a water \v{C}erenkov detector, the event sample is restricted to 
charged-current (CC) ``quasi-elastic-like''  (QE-like) events, a 
definition which was first introduced in 
Refs.~\cite{Delorme:1985ps,Marteau:1999jp,Martini:2009uj}. 
These are selected by requiring that there be only one charged 
particle above \v{C}erenkov threshold in the final state, the
so-called ``single ring'' events.  For a QE event considering the 
neutron at rest, the neutrino energy can be reconstructed from the 
kinematic variables of the charged lepton $\ell$ in the final state as:
\begin{align}
\nonumber &E^{QE}_{\nu} = \\
\nonumber &\frac{2(M_{n}-E_{b})E_{\ell}-(E^{2}_{b}-2M_{n}E_{b}+\Delta M^{2})}{2(M_{n}-E_{b}-E_{\ell}+p_{\ell}\cos\theta_{\ell})},\\
\label{eq:recengwoproton}
\end{align}
where $M_{n}$ is the free neutron rest-mass, $\Delta M^2 =
M_n^2-M_p^2+m^2_\ell$, and $E_{b}$ is the binding energy. 
Both GENIE and GiBUU use $E_{b}$=30~MeV, a value which is
obtained from electron scattering data~\cite{Moniz:1971mt,VanOrden:1980tg}. 
Eq.~\ref{eq:recengwoproton} is exact only for the QE 
interaction with a neutron at rest. However, for any neutrino 
experiment observing QE-like events, the final sample will also 
contain events that are not QE. For example, in a single pion 
neutrino interaction, \textit{i.e.}, an event with a charged lepton 
and a pion in the final state, the pion can be absorbed
by the nucleus during final state interactions (FSI). Such event will
therefore be classified as QE-like. In this case, if Eq.~\ref{eq:recengwoproton} 
is used to reconstruct the neutrino energy, it will unavoidably lead 
to a value of the reconstructed neutrino energy \emph{lower} than 
the true incident value for the reason that part of the neutrino
energy is carried away by the pion and eventually absorbed by the 
nucleus. The actual value of the reconstructed neutrino energy will 
depend on the energy of unobserved particles in the final
state~\cite{Leitner:2010kp}. Therefore, while the
reconstructed energy will mostly coincide with the true energy of the
incident neutrino for a true QE event, there exists a certain
probability to have a non-QE event end up being reconstructed 
with a significantly different neutrino energy. This defines a
migration matrix between true and reconstructed neutrino energies, 
$N(E^{\rm rec}$,$E^{\rm  true})$, where each element represents the
probability that an event for each given true neutrino energy 
$E^{\rm true}$ ends up being reconstructed with a different energy 
$E^{\rm rec}$.
The final QE-like event sample will comprise QE as well as non-QE
events, where pions are not present in the final state:
\begin{widetext}
\begin{eqnarray}
\nonumber && N^{QE-like}_i =  \sum_j M^{QE}_{ij} N^{QE}_j + \sum_{non-QE}\sum_j M^{non-QE}_{ij} N^{non-QE}_j  \\
&\propto & \sum_j M^{QE}_{ij} \phi_\alpha(E_j)P_{\alpha\beta}(E_j)\sigma^{QE}_\beta(E_j) + \sum_{non-QE} \sum_j M^{non-QE}_{ij} \phi_\alpha(E_j)P_{\alpha\beta}(E_j)\sigma^{non-QE}_{\beta,0\pi}(E_j) \, ,\nonumber \\
\label{eq:events-QElike}
\end{eqnarray}
\end{widetext}
where $E_j$ is the true neutrino energy, $P_{\alpha\beta}$ stands for
the oscillation probability of $\nu_\alpha \rightarrow \nu_\beta$, 
$\phi(E_j)$ is the initial flavor neutrino flux, the matrices 
$M_{ij}\equiv N(E^{rec}_i, E^{true}_j)$ account for the probability
that an event with a true neutrino energy in the bin
$j$ ends up being reconstructed in the energy bin $i$.
$\sigma^{non-QE}_{0\pi}$ stands for the cross-section for a given
non-QE process in which there are no pions in the final state. In this work, 
we will study a muon neutrino disappearance experiment, so $\alpha=\beta=\mu$.
\par Finally, it should be kept in mind the classification of event types
in neutrino experiments is not well-defined, since the incident
neutrino energy is not known. In this work, we focus on the migration
of non-QE events into the QE-like sample. However, it should be kept
in mind that there is a second source of mis-identifying events
which takes place in the opposite direction. For instance,
an initially purely QE interaction, where the outgoing proton
re-interacts inside the nucleus, producing a Delta resonance
($\Delta$). The $\Delta$ will then decay and produce a pion in the 
final state. As a result, this event will be classified as non-QE due
to the resonance production. In summary, both the QE and resonance 
interactions are entangled, and sometimes it is hard to distinguish 
one from the other. Neutrino oscillation experiments rely on different 
event generators to help them estimate the portion of mis-identified 
QE and resonance production events. However, there are considerable
theoretical uncertainties in predicting both event classes. This
results in different event generators having different theoretical
implementations of the same event types. This is a major source of
systematic uncertainty which, like all theory-related errors, is
difficult to quantify.

\section{Event generators}
\label{sec:gen}

In this work, both migration matrices and cross-sections in
Eq.~\ref{eq:events-QElike} are computed by the event generator. We
have considered two different event generators in this work: GENIE
(Generates Events for Neutrino Interaction Experiments)
2.8.0~\cite{Andreopoulos:2009rq} and GiBUU (Giessen
Boltzmann-Uehling-Uhlenbeck) 2.6~\cite{Leitner:2009ke,Buss:2011mx}. 
GENIE is used by the major neutrino accelerator experiments in the US, 
such as MINER$\nu$A~\cite{Drakoulakos:2004gn}, MINOS~\cite{Adamson:2007gu},
MicroBooNE~\cite{Chen:2007ae}, NO$\nu$A\cite{Ayres:2004js} and LBNE~\cite{CDR,Adams:2013qkq},
and is also used by the T2K~\cite{Itow:2002rk} experiment~\footnote{T2K also uses an independent event generator, NEUT~\cite{Hayato:2009zz}.}.  
GiBUU, on the other hand, is based on a semiclassical transport
model~\cite{Buss:2011mx}, and therefore constitutes a complimentary 
and independent theoretical approach.
\par
\begin{figure}
\begin{center}
\subfigure[~Cross-sections computed with GENIE]{\includegraphics[width=1.0\columnwidth]{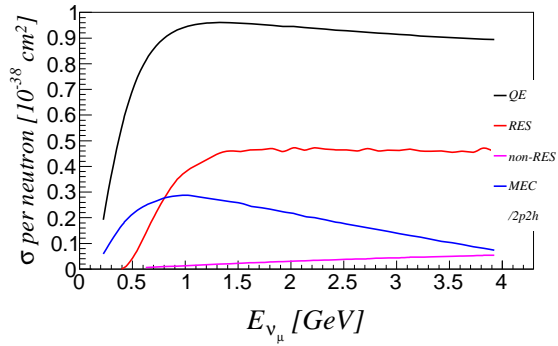}\label{fig:xsec_genie}}
\subfigure[~Cross-sections computed with GiBUU]{\includegraphics[width=1.0\columnwidth]{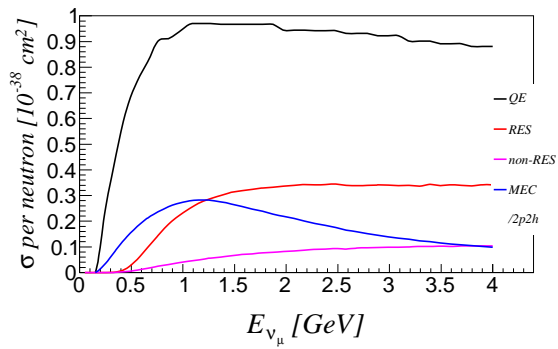}\label{fig:xsec_gibuu}}
\caption{(color online) \textit{Total cross-sections per neutron as a function of the
    true neutrino energy for different charged-current processes on $^{16}$O after
    requiring no pion in the final state. Results are shown for the two event generators 
    considered in this work. Note: RES here only represents the $\Delta$-resonant 
    pion production without including the heavier resonance modes into consideration. 
    Non-RES in panel (a/GENIE) includes the exclusive
    coherent pion production, while in panel (b/GiBUU) includes only the
    incoherent part.}}
\label{fig:tot_xsec_genie_gibuu}
\end{center}
\end{figure}

 Figure~\ref{fig:tot_xsec_genie_gibuu} 
shows the total cross-sections per neutron on $^{16}$O with no pions in the final state 
for all QE-like interactions using GENIE and GiBUU. 
In this work, we consider charged-current quasi-elastic (QE),
charged-current single pion production (RES) from $\Delta$ resonant
decay, charged-current non-resonant pion production (non-RES), and 
neutrino interactions involving the two-nucleon currents arising from 
meson exchange processes and nucleon-nucleon correlations (MEC/2p2h). 
We analyze simulated events with the additional requirement of no 
pion in the final-state. It should be noted that we do not consider 
the contributions of deep inelastic scattering (DIS) and the
production of higher resonances, even though they were included 
in the analysis performed in Ref.~\cite{Coloma:2013rqa}. The reason for this is the following. 
The cross-section for the
production of higher resonances with no pion in the final state is 
found to be very small, see Fig.~1 in Ref.~\cite{Coloma:2013rqa}, 
and so is its contribution to the total number of events. In addition, 
for the particular setup in this simulation work, the neutrino flux 
decreases very rapidly above 1-1.5~GeV, where the DIS cross-section 
is still negligible. Therefore, since there is no sizable contribution from 
DIS events either, they have been removed from our simulations. 
It should be kept in mind, though, that the removal of events coming
from the higher resonances and DIS may not be the case for neutrino 
experiments in which the flux peaks at higher energies, such as 
LBNE~\cite{CDR,Adams:2013qkq}.
%
%
\par
The migration matrices are computed as follows. Each neutrino
interaction sample consists of 50,000 events, generated in bins of
true neutrino energy from 0.2~GeV to 4.8~GeV, for a total of 46
bins. Each matrix is built by counting the number of entries
in each bin of the reconstruction energy as defined in
Eq.~\ref{eq:recengwoproton} and then dividing it by the total amount
of entries. This is done separately for each type of interaction 
as defined in Fig.~\ref{fig:tot_xsec_genie_gibuu}. 
For each given value of the initial true energy, a probability distribution 
in reconstructed energies $\rho^{E}_{\rm rec}$ can be computed. 
In fact, $N(E^{\rm rec}$,$E^{\rm true})$ can be regarded as filling one 
two-dimensional probability density plane where the y-axis 
indicates the density in reconstructed energies, $\rho^{E}_{\rm rec}$, 
and the x-axis represents the true energy value. This effectively
implements an energy smearing due to nuclear effects. We require 
the sum of densities for a given true energy to be normalized to the 
unity, so that the number of events before and after migration remains 
the same.
\par Both GENIE and GiBUU event generators differ on the nuclear
models used as well as on how the different types of interactions are computed. 
In the rest of this section, we list some of the main differences between them.

%
\subsection{Nuclear Model}
When generating neutrino-nucleus QE interactions, both GENIE and GiBUU use variants of the relativistic
Fermi gas model (RFG)~\cite{Smith:1972xh} to describe the 
nuclear structure and the dynamics of neutrino-nucleus interactions 
under the hypothesis of plane wave impulse approximation (PWIA), see for instance Ref.~\cite{Benhar:2006wy} for a review. 
In the PWIA, the struck nucleon in every single neutrino-nucleus interaction
is treated as a quasi-free particle due to the high momentum transfer Q$^2$, 
while the rest of the nucleus, the so-called spectator system, is left unperturbed. 
In the RFG, the double differential cross-section can be written down 
as a function of scattering angle and outgoing lepton's total energy,
that is approximated as the lepton's kinematic energy~\cite{Benhar:2005dj}:
\begin{equation}
\label{eq:tot_xsec_formula}
\frac{d^{2}\sigma_{\rm IA}}{d\Omega\, dE_{\mathit l}} = \int{d^{3}p~dE~P_{RFG}({\textbf p},E)\frac{d^{2}\sigma_{elem}}{d\Omega\,dE_{l}}}, \\
\end{equation}
with
\begin{equation}
\nonumber P_{RFG} ({\textbf p},E) = \frac{6\pi^{2}A}{p^{3}_{F}}\theta({\rm p_F}-{\textbf p})\delta(\Delta E), \\
\end{equation}
where $6\pi^{2}A/p^{3}_{F}$ is a normalization factor, and $p_F$ is the
Fermi momentum ($221\,{\rm MeV}$, or $1.12\,{\rm fm^{-1}}$), 
which is the same for all nuclear targets in GENIE and GiBUU. $\Delta E = (E_p-E_b+E)$, 
where $E_{b}$ is the average binding energies (\textit{i.e.}, the
binding energy of the nucleon). $\sigma_{\rm elem}$ is the elementary 
cross-section used to describe the probability of interactions between 
the incident neutrino and the nucleon. The integration limits in 
Eq.~\ref{eq:tot_xsec_formula} are determined by the boundaries of 
the kinematically allowed region~\cite{Benhar:2006nr}.
\subsection{Quasi-elastic Scattering}
The nuclear model used in GENIE to simulate QE interactions is a
modified RFG that includes short-range nucleon-nucleon correlations
according to the model developed by A. Bodek and
J. L. Ritchie~\cite{Bodek:1981wr}. GiBUU also uses the RFG, but in
this case, the RFG is modified by adding corrections from the nucleon's
momentum and density dependent mean-field potential, where all
nucleons are considered to be bound. The phase-space density function 
in GiBUU also includes the real part of the self-energy for the
knock-out nucleon~\cite{Leitner:2008ue}. Both generators use the same
value for the axial mass, $\rm M_{\rm A}=$1~GeV/c$^{2}$.  
\par
\begin{figure}[t!]
\includegraphics[width=1\columnwidth]{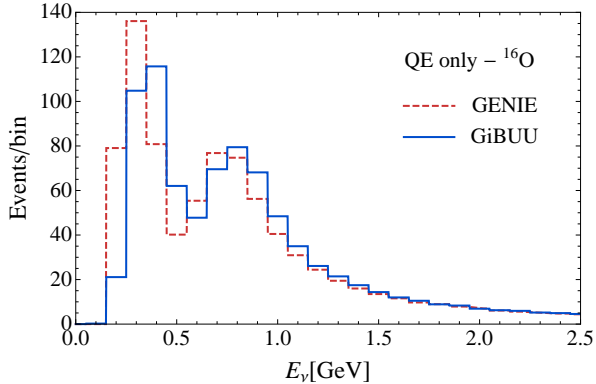}
\caption{(color online)~\textit{Charge Current Quasi-Elastic event distributions as a function of the reconstructed neutrino energy for both GENIE (dotted red lines) and GiBUU (solid blue lines). In both cases, oxygen is used as the target nucleus to obtain the cross sections. Both curves include final state interactions, and detector efficiencies have been accounted for. }}
    \label{fig:events_qe}
\end{figure}

 The vector form factor in GENIE is BBBA05~\cite{Bradford:2006yz},
while in GiBUU it is BBBA07~\cite{Bodek:2007ym}. The QE cross-sections 
for the two generators are practically the same but with small discrepancies 
which result from different electromagnetic form factor shapes and
separate corrections added to the RFG. The
QE event distribution for GENIE and GiBUU on $^{16}$O is shown in 
Fig.~\ref{fig:events_qe}.
%
%
\par 
We have noted some differences in the event distribution as function
of reconstructed neutrino energy between GENIE and GiBUU as shown 
in Fig.~\ref{fig:events_qe}. The nuclear model used to describe the QE neutrino interactions is essentially the same between GENIE and GiBUU. For example, we find no differences in the QE cross-section as shown in Fig.~\ref{fig:tot_xsec_genie_gibuu} between GENIE and GiBUU. We do not find any shift between the event distributions as function of reconstructed neutrino energy if we repeat the same simulation but removing FSI effects for both GENIE and GiBUU. Nevertheless, we find a shift of 10\% between the event distribution as a function of reconstructed neutrino energy (defined as in Eq.~\ref{eq:recengwoproton}) between GENIE and GiBUU, as it is shown in Fig.~\ref{fig:events_qe}. This is not surprising due to the fact that GENIE and GiBUU follow a completely different approach to describe FSI. For example GiBUU, the FSI is modeled by solving the semi-classical Boltzmann-Uehling-Uhlenbeck equation. For more details on GENIE and GiBUU, see for instance Refs.~\cite{Leitner:2006ww,Leitner:2009ke,Andreopoulos:2009rq}.
\par The migration matrices for QE events are shown in
Fig.~\ref{fig:genie_matrix_qe} and Fig.~\ref{fig:gibuu_matrix_qe} 
in App.~\ref{app:matrices} for GENIE and GiBUU, respectively. These 
matrices include FSI effects.
%
%
%
\subsection{Meson Production via Baryon Resonances}
\label{subsec:gen-nonres}
At the neutrino energies relevant for this work ($\sim$1~GeV), the
second dominant neutrino interaction is the single pion production 
via $\Delta$ resonances (RES). 
The RES includes: (1) the baryon resonance decay; 
(2) the charge-exchange of a neutral pion inside the nucleus; 
(3) the absorption of multiple pions.
\par  GiBUU contains 13 kinds of resonance modes. The vector form
factors for each resonance mode are obtained from the
MAID~\cite{Drechsel:1992pn, Drechsel:1998hk} analysis of the electron
scattering data. The axial form factors for GiBBU are derived from a fit to the
data as described in Ref.~\cite{AlvarezRuso:1998hi}. In the case of GiBUU, nuclear
effects include collisions within nucleons and the nucleon's momentum
and density dependent mean-field potential. The FSI is determined by 
modeling a coupled-channel transport method. This model is tested 
against electron-, photon-, pion- and proton-scattering data.
\par GENIE, instead, applies the Rein-Sehgal~\cite{Rein:1981ys} model with
16 resonance models in the region of transferred energy ${\rm W}\, <\, {\rm W}_{\rm cut} =
1.7\,{\rm GeV}$. Fermi motion and Pauli blocking are the only nuclear
effects included in this case. The FSI is modeled using an intra-nuclear model
as described in detail in Ref.~\cite{Dytman:2009zz}. GENIE RES is validated with electron
scattering data from $^{56}$Fe and $^{12}$C targets.  The RES
event distributions for GENIE and GiBUU using $^{16}$O are shown in
Fig.~\ref{fig:events_res} in App.~\ref{app:matrices}. The
corresponding migration matrices are shown in Figs.~\ref{fig:genie_matrix_res} 
and~\ref{fig:gibuu_matrix_res} for GENIE and GiBUU, respectively.
In Fig.~\ref{fig:tot_xsec_genie_gibuu}, we only show the contribution
of $\Delta$-resonant production to the cross-section without regarding
the heavier resonance modes. As already explained at the beginning of this section, the
higher resonance effects become more significant and dominant 
at higher neutrino beam energies and would therefore have little impact on our results.
\subsection{Two Particles - Two Holes and Meson Exchange Currents}
Aside from QE and RES, at neutrino energies below 1~GeV an additional 
contribution to the total neutrino cross-section
comes from processes involving two particle-two hole (MEC/2p2h) excitations, as shown in Fig.~\ref{fig:tot_xsec_genie_gibuu}.
Several models have been proposed to compute these contributions in neutrino experiments, and the field has been object of a very intense research in the past few years. For an incomplete list of references on this topic, see~\cite{Martini:2009uj,Martini:2010ex,Benhar:2010nx,Amaro:2010sd,Juszczak:2010ve,
Bodek:2011ps,Nieves:2011pp,Nieves:2011yp,Martini:2011wp,Lalakulich:2012ac,Benhar:2013bba,Benhar:2013bwa,Gran:2013kda}. 
Sources of MEC/2p2h excitation include:
(1) nucleon-nucleon correlations in the initial state, (2) neutrino coupling to 
2p2h and (3) FSI. In processes in which two-nucleons are knocked out
from the target nucleus, the nucleon's momentum distribution in the
spectator system is influenced. Therefore, an excess of QE-inclusive 
events is produced. Furthermore, it is obvious that if Eq.~\ref{eq:recengwoproton} is applied to obtain the reconstructed neutrino energy, this will most likely differ from the true incident neutrino energy, due to the non-QE nature of the interaction. The effect of 2p-2h has been recently revealed by 
the theoretical interpretation~\cite{Martini:2009uj,Nieves:2011pp} of the results obtained in the MiniBooNE 
experiment~\cite{AguilarArevalo:2010zc}.
\par
A detailed description of the implementation of MEC/2p2h in GENIE is
available in Ref.~\cite{Katori:2013eoa}, and for GiBUU is available in Ref.~\cite{Lalakulich:2012ac}.
\par The MEC/2p2h event distribution for GENIE and GiBUU on $^{16}$O 
is shown in Fig.~\ref{fig:events_2p2h} of App.~\ref{app:matrices}. The 
associated migration matrices are shown in
Figs.~\ref{fig:genie_matrix_2p2h} and~\ref{fig:gibuu_matrix_2p2h}
for GENIE and GiBUU, respectively. These matrices seem to be rather different. 
To simulate this particular interaction, both GIBUU and GENIE have been tuned based on the 
measurements on $^{12}$C done by the MiniBooNE experiment and they do not yet include full theoretical model implementations like the ones presented in Refs.~\cite{Martini:2009uj,Martini:2010ex,Nieves:2011pp,Nieves:2011yp}. In addition, it should be noted that the MiniBooNE experiment measures the sum of the QE and MEC/2p2h contributions, but the experiment is not capable of discriminating between them. In other words, the tuning of the neutrino interaction generators is performed in such a way that the sum of the QE and the MEC/2p2h contributions to the double differential cross section agrees with the data. Therefore, if a difference exists between GENIE and GiBUU in the QE interactions, then the MEC/2p2h by construction will be different as well. Finally, it is worth mentioning that no difference in the total cross-section between GENIE and GiBUU is found for both $^{12}$C and $^{16}$O as a result of the heavy tuning of the model on MiniBooNE $^{12}$C results.
%
\subsection{Non-Resonant Pion Production}
The non-resonant pion production (non-RES) includes the contributions
of coherent and non-coherent pion productions. The coherent pion production 
results in one muon and one single pion in the
final state, in which the pion is not produced by a $\Delta$-resonance decay.
The residual nucleus system remains in its ground state in the coherent production while in the incoherent pion 
production this is not the case. GENIE includes only the coherent (exclusive) pion 
production~\cite{Boyd:2009zz}, whereas GiBUU considers only the incoherent
pion production process. A full theoretical calculation for this
interaction mode can be found in Ref.~\cite{Nakamura:2009iq,Nakamura:2011rt}. 
\par GENIE uses PCAC~\cite{Boyd:2009zz} with the
Rein-Sehgal~\cite{Rein:1982pf} model to simulate the single pion 
interaction with the nucleon inside the ground-state target nucleus. 
The neutrino-nucleon cross-section is computed at the initial time
(t=0) with the elastic nuclear form factors and an absorption factor
to simulate the FSI of the outgoing pion. In contrast, GiBUU takes 
into account only the incoherent part of the initial-state pion, 
and does not make any local approximation to the $\Delta$ propagator 
in the medium~\cite{Leitner:2009ph,Boyd:2009zz}. 
The non-RES event distribution for GENIE and GiBUU on $^{16}$O is
shown in Fig.~\ref{fig:events_coh} of App.~\ref{app:matrices}. The
associated migration matrices are shown in Figs.~\ref{fig:genie_matrix_coh} 
and~\ref{fig:gibuu_matrix_coh} for GENIE and GiBUU, respectively.

\section{Simulation details}
\label{sec:sim}
%
In this work we consider a setup very similar to the T2K experiment,
simulated following Ref.~\cite{Huber:2009cw}. The main details of the
setup are summarized in Tab.~\ref{tab:setup}. We consider two
detectors: a far detector, placed at 295~km from the source, with a
fiducial mass of 22.5~kton; and a near detector of 1~kton fiducial
mass, placed at 1~km from the decay pipe. The size and location of the
near detector have been chosen so as to guarantee that it observes
enough events to be able to constrain the systematic errors included
in the analysis (for details on the systematic errors and the $\chi^2$
implementation, see App.~\ref{app:chi2}). In this work, we assume that
the two detectors observe the same flux, and that they are
identical in terms of their composition and detection properties. It
should be kept in mind, however, that these assumptions would most
likely \emph{not} be realized in an actual neutrino beam
experiment. This may lead to a \emph{larger} impact of nuclear effects
on the extraction of the oscillation parameters than what is found in
this work.
%
\begin{table*}[htbn!]
\centering
\begin{tabular}{l@{\quad}c@{\quad}c@{\quad}c@{\quad}c@{\quad}c@{\quad}} 
\hline \hline
 & Baseline  & Fiducial mass & Flux peak & Beam Power & Running time \\ \hline
Far     &  295~km    & 22.5~kt  & \multirow{2}{*}{0.6~GeV}  &  \multirow{2}{*}{750~kW}  & \multirow{2}{*}{5~yrs}  \cr  
Near       &  1.0~km     & 1.0~kt    & 			&		  &			 \\ \hline\hline 
\end{tabular}
\caption{\textit{Main details for the experimental setup simulated in this
  work.} \label{tab:setup}}
\end{table*}
%
\par The only oscillation channel considered is $\nu_\mu \rightarrow
\nu_\mu$ disappearance. For this channel, the only relevant background
would come from neutral current (NC) interactions. We do include such
a background in our analysis, which produces a total of $\sim 275$ events at
the far detector. Nevertheless, since our aim is to explore the impact
of nuclear effects on the CC signal, we keep the background event
rates the same for all the configurations under consideration in this
work, and no study is done on the variation of these rates with
different nuclear models and/or energy reconstruction
effects. Background rates are smeared using a gaussian with 
$\sigma(E)=85$~MeV, following Ref.~\cite{Huber:2009cw}. For the
signal, on the other hand, since the migration matrices due to nuclear
effects already introduce a rather coarse energy smearing, we do not
consider additional effects due to the finite resolution from the
detector. We believe the effect due to this will be certainly minor,
considering the large smearing that we observe already for the
QE event sample, see Figs.~\ref{fig:genie_matrix_qe}
and~\ref{fig:gibuu_matrix_qe} in App.~\ref{app:matrices}. Finally, energy dependent detection
efficiencies are implemented for the signal, following
Ref.~\cite{Huber:2009cw}. These are applied after the events are
migrated to reconstructed neutrino energies.
\par
\begin{figure*}[htbn!]
\subfigure[~Expected events at the far detector]{\includegraphics[width=1.0\columnwidth]{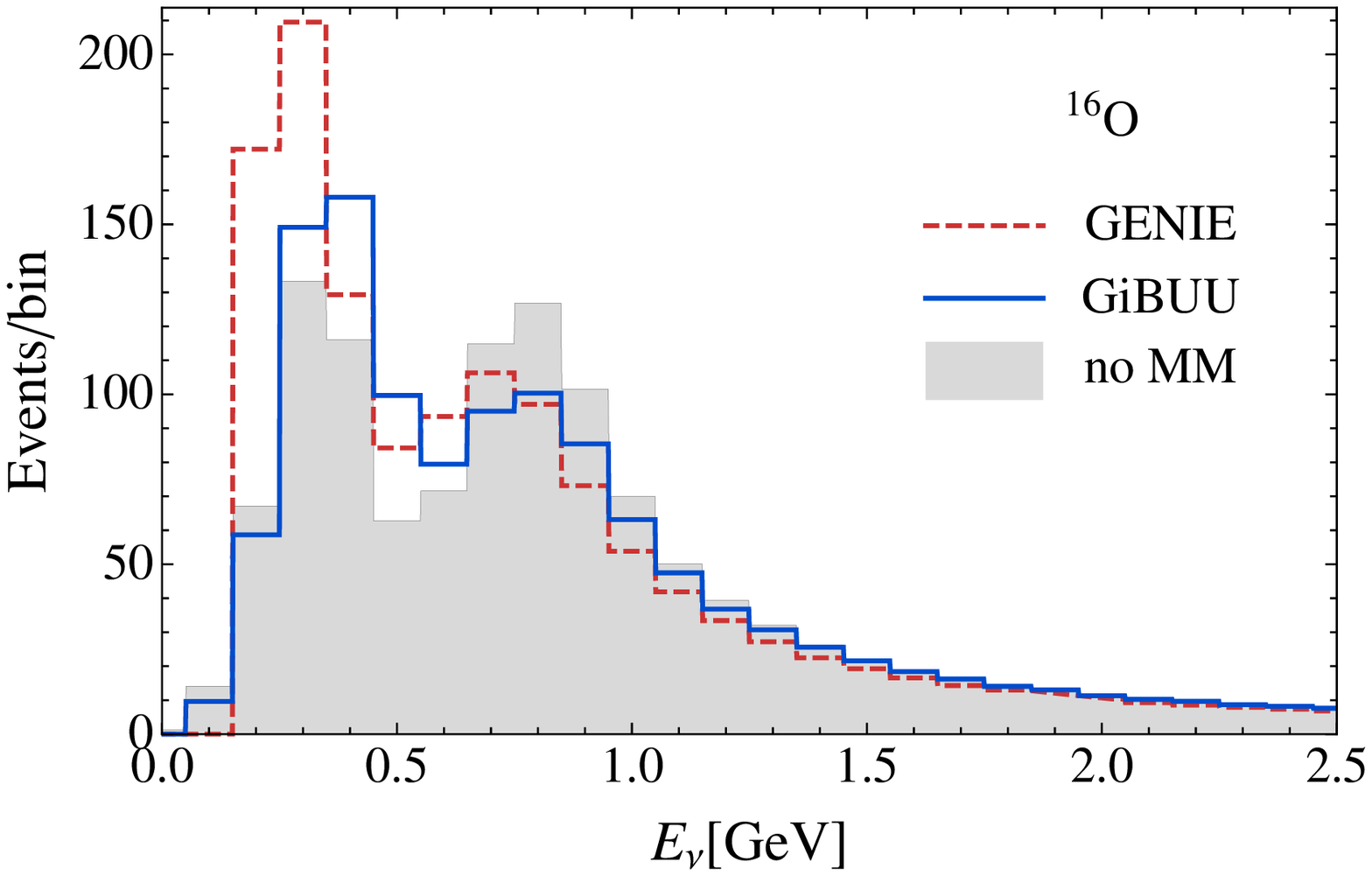}}
\subfigure[~Expected events at the near detector]{\includegraphics[width=1.0\columnwidth]{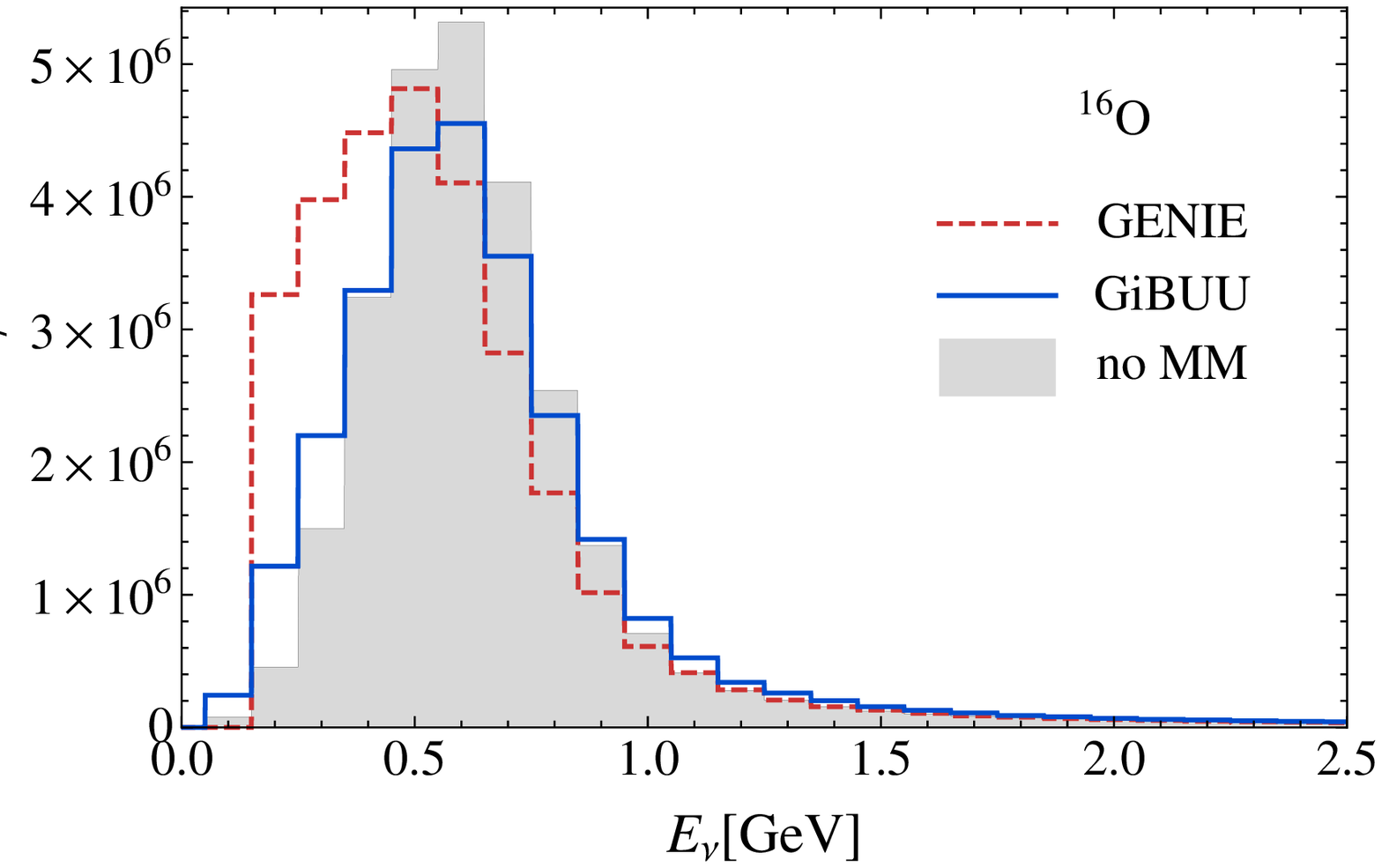}}
\caption{(color online)\textit{ Binned QE-like event rates as a
    function of the reconstructed neutrino energy in GeV, computed
    using Eq.~\ref{eq:events-QElike}. The solid blue (dashed red)
    lines show the event rates obtained after migration using the
    GiBUU (GENIE) event generators. The shaded areas show the expected
    event rates coming from the QE-like event sample computed using
    the GiBUU cross-section for $^{16}$O, as for the solid blue lines,
    but without including any migration matrices (\textit{i.e.},
    taking $M^{QE}_{ij}=M^{non-QE}_{ij}=\delta_{ij}$ in
    Eq.~\ref{eq:events-QElike}). For the shaded areas, a gaussian
    energy resolution function with a constant standard deviation of 85~MeV is added to account for the finite resolution of
    the detector. Left and right panels show the event rates at the
    near and far detectors, respectively. All lines have been obtained
    for the oscillation parameters in Eq.~\ref{eq:oscparams}, and
    detector efficiencies have already been accounted for.  }}
\label{fig:rates}
\end{figure*}

 The expected number of events at the far detector for the different
contributions to the QE-like event sample are shown in
Tab.~\ref{tab:events}. These are computed using the cross-sections produced for the two event generators under
consideration, using oxygen as the target
nucleus. Fig.~\ref{fig:rates} shows the expected event rates for the
QE-like event sample, binned as a function of the neutrino
energy. Results are shown for the near and far detectors in the right
and left panel, respectively. The gray shaded areas show the event
rates for the events before migration to reconstructed energies, while
the solid blue (dashed red) lines show the results after migration to
reconstructed neutrino energies, when the matrices and cross-sections are computed using
GiBUU (GENIE), as explained in Sec.~\ref{sec:gen}. In all cases, the
oscillation parameters have been set to the values in
Eq.~\ref{eq:oscparams} in App.~\ref{app:chi2}, and detector efficiencies are accounted for. We find that the event distributions using the matrices generated with GiBUU are in agreement with those shown in Fig.~10 of Ref.~\cite{Lalakulich:2012hs}. Similar results were also shown in Fig.~3 of Ref.~\cite{Martini:2012uc} for a different nuclear model\footnote{It should be noted that, in Fig.~\ref{fig:rates},  the event distribution before migration has been smeared with a gaussian energy resolution function with a constant width, to account for the detector�s finite energy resolution. As a result, the lowest two energy bins have slightly more events before migration than after migration. We have checked that for a detector of perfect energy resolution this is not the case and therefore our results are in agreement with those in Refs.~\cite{Lalakulich:2012hs,Martini:2012uc}. }. 
%
%
\begin{table*}[htbn!]
\renewcommand{\arraystretch}{1.2}
\begin{tabular}{l@{\quad}c@{\quad}c@{\quad}c@{\quad}c@{\quad}c@{\quad}} 
\hline \hline
             & QE & RES & non-RES & MEC/2p2h & Total\\ \hline 
GiBUU        & 870 & 152 & 32 & 214 & 1268 \\ 
GENIE       & 877 & 221 & 11 & 249 & 1358 \\
\hline\hline
\end{tabular}
\caption{\textit{Total number of events expected at the far detector, for the
  different contributions to the QE-like sample, and for the
  oscillation parameters in Eq.~\ref{eq:oscparams}. The expected
  number of events are shown for the two event generators under
  consideration. In both cases oxygen is chosen as the target
  nucleus. Efficiencies are already accounted for. The distribution of
  events for the different contributions as a function of the
  reconstructed neutrino energy can be found in
  App.~\ref{app:matrices}.} \label{tab:events}}
\end{table*}
%
\par It is also worth mentioning the large contribution to the QE-like
sample coming from MEC/2p2h contributions. It has recently been argued~\cite{Martini:2009uj,Nieves:2011pp} that this 
may be the source of the large discrepancy between the value of $M_A$
reported in Ref.~\cite{AguilarArevalo:2010zc} and the world average value of
$M_A\sim1.0$~GeV, see for instance Ref.~\cite{Bernard:2001rs}. 
As it can be seen from Tab.~\ref{tab:events} (see also Fig.~\ref{fig:events_2p2h} in
App.~\ref{app:matrices}), the expected contribution to the QE-like
sample from MEC/2p2h interactions is rather large for the setup considered
in this work as well. Table~\ref{tab:events} shows that around a $\sim
17\%$ of the final QE-like sample in our simulated setup would come
from MEC/2p2h interactions. It is also noticeable the difference between
the number of events ($\sim 10\%$) obtained when the cross-section is computed using
GiBUU or GENIE. This
difference is not coming from the size of the cross-section, since in
both cases the MEC/2p2h cross-section is tuned to the MiniBooNE data and
the size of the peak is roughly the same, see
Fig.~\ref{fig:tot_xsec_genie_gibuu}. The reason for the difference in
the number of events is a shift in neutrino energy in the cross
section obtained from the two generators when they are compared
against each other: while the GENIE cross-section reaches its maximum
at around 1~GeV, the GiBUU cross-section peaks at slightly higher
energies, around 1.2~GeV. Since the peak in the neutrino flux
considered in this work lies below 1~GeV or so, the number of events
from MEC/2p2h interactions when the GENIE cross-section is used will be
larger. Finally, small differences can also be appreciated in the number of events coming from resonant and non-resonant pion production. The first one comes from a different number of resonances included in the computation, see also Sec.~\ref{subsec:gen-nonres}. In the second case, a different result is obtained since GiBUU computes incoherent pion production while GENIE computes only the coherent contribution to this process. 

\section{Results}
\label{sec:results}

In this section, we explore the impact on the extraction of the
oscillation parameters in three different scenarios:
\begin{enumerate}[(A)]
\item when the target nucleus is changed in the fit;
\item when the nuclear model is changed in the fit;
\item when multi-nucleon contributions (MEC/2p2h) are completely removed from the fit.
\end{enumerate}

\subsection{Impact of different target nuclei}
\begin{figure*}[htbn!]
\subfigure[~QE-like event distributions ]{\includegraphics[width=1.0\columnwidth]{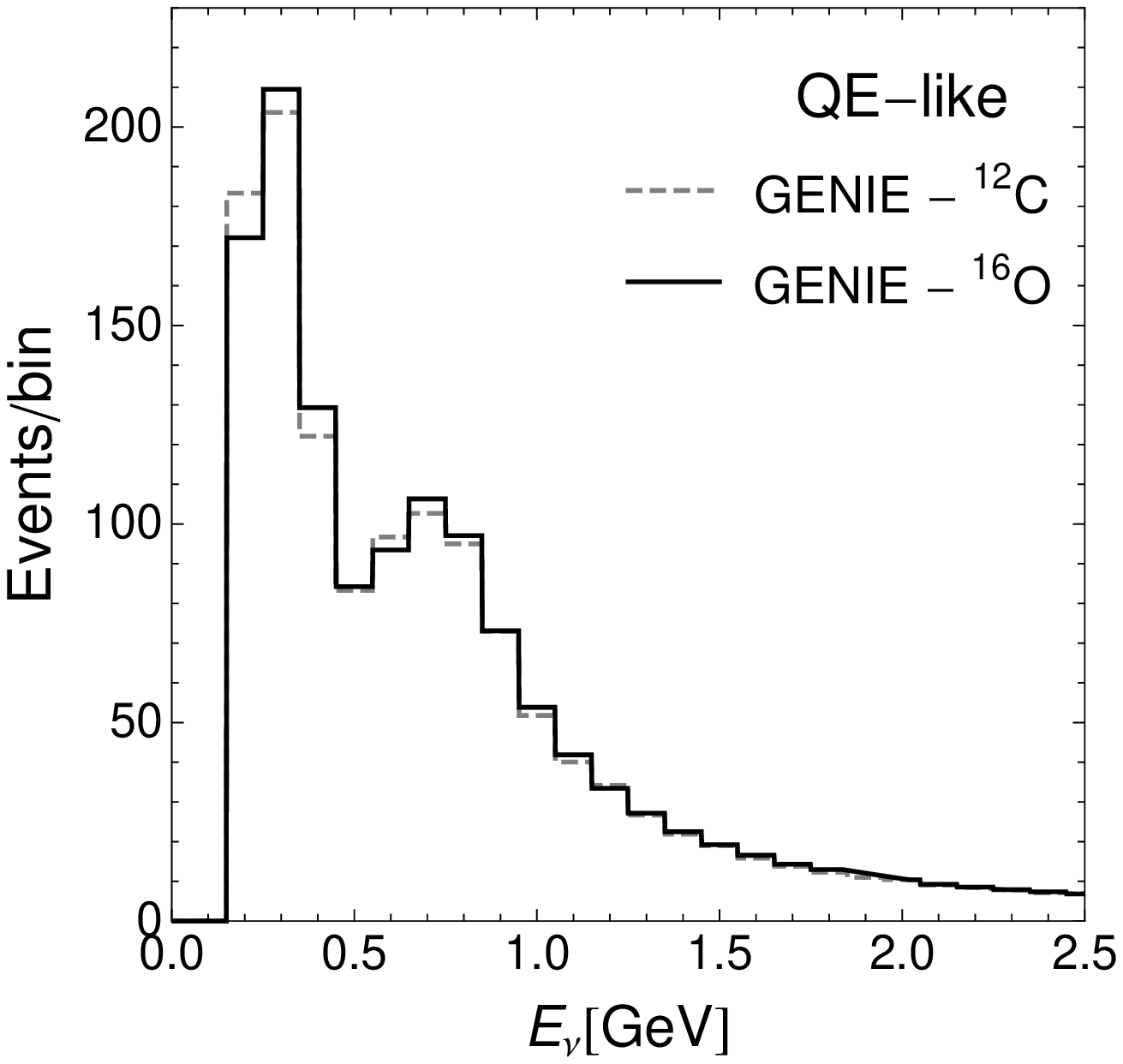}\label{fig:targets1}}
\subfigure[~Confidence regions ]{\includegraphics[width=1.0\columnwidth]{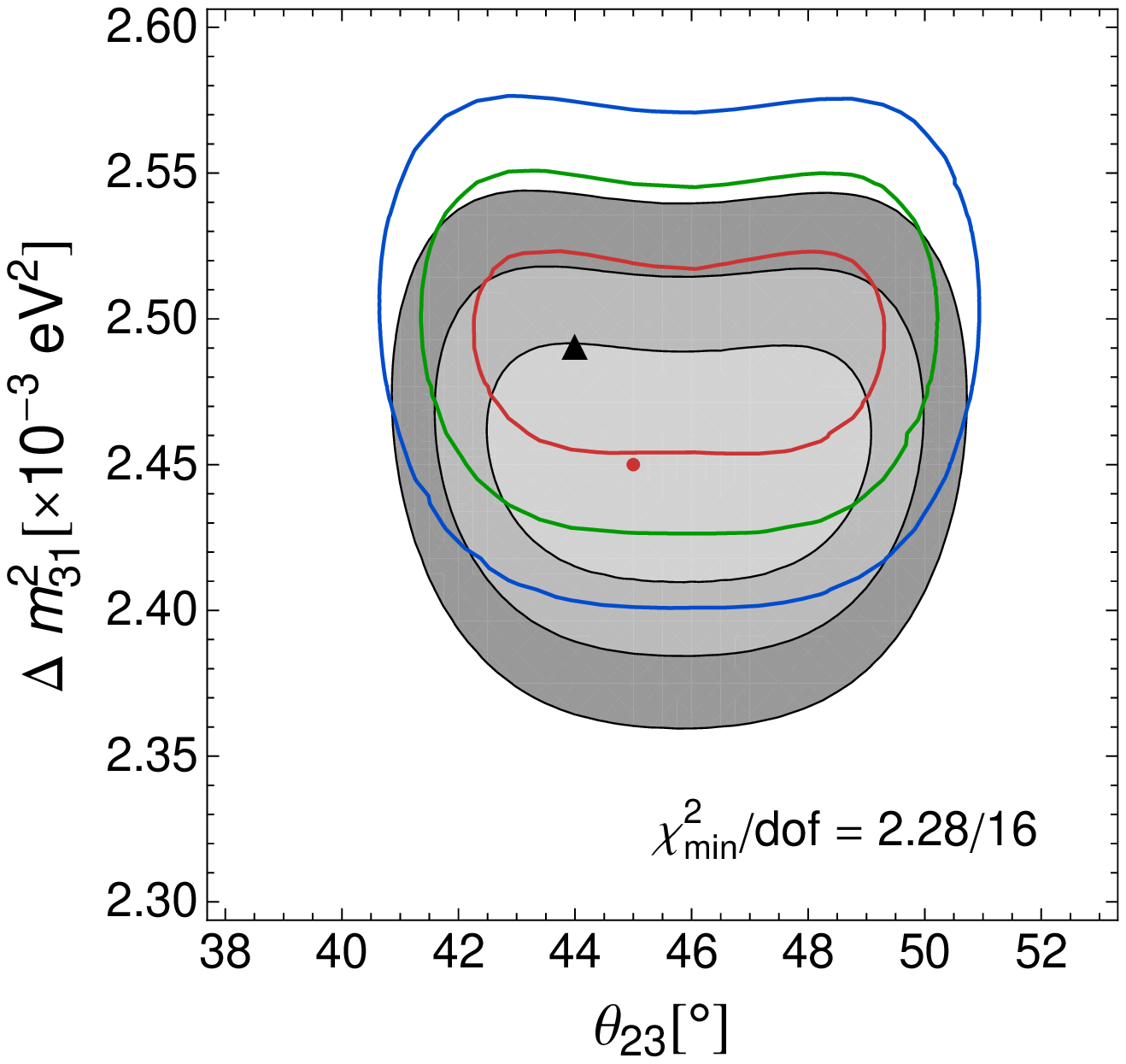}\label{fig:targets2}}
\caption{(color online)\textit{ Impact on the results if a different
    target is used to compute the true and fitted rates in the
    analysis. Left: Event rates per bin obtained using migration
    matrices computed for oxygen (solid black) or carbon (dashed gray)
    as the target nucleus. Right: Result of the fit in the in the
    $\theta_{23}-\Delta m_{31}^2$ plane. The shaded area shows the
    confidence regions that would be obtained at 1, 2 and 3$\sigma$ 
    if the true and fitted rates are generated using the same
    set of migration matrices. The colored lines show the resulting
    regions if the event rates computed using matrices for oxygen
    (solid lines in the left panel) are fitted with the rates computed
    using matrices for carbon (dashed lines in the left panel). The
    GENIE event generator has been used in both cases to generate the
    migration matrices. The red dot indicates the true input value,
    while the black triangle shows the location of the best fit
    point. The value of the $\chi^2$ at the best fit is also shown, together with the number of degrees of freedom.}}
\label{fig:targets}
\end{figure*}

It is common practice to ``tune'' event generators according to a
given target nucleus in a certain experiment. However, the event
generator may be used later on for an oscillation experiment using a
different target. In this section we evaluate the effect on the
oscillation analysis if an event generator is tuned according to data
obtained for a certain target but the experiment is performed using a
different target. Results are shown in Fig.~\ref{fig:targets}. In the
left panel, the binned expected event rates are shown as a function of
the reconstructed neutrino energy, when oxygen (solid black lines) and
carbon (dashed gray lines) are used to compute the cross-sections and migration matrices
in Eq.~\ref{eq:events-QElike}. In both cases, the GENIE event
generator is used to compute the matrices for all contributions. As it
can be seen from the figure, there is a slight shift towards lower
energies in the event rates for carbon with respect to those obtained
for oxygen. We have checked that this shift is already present for the true QE event sample, 
and that it remains if final state interactions are removed from the simulation. 
This is automatically translated into a shift for the best
fit value of the mass-squared splitting. The effect is shown in the
right panel, where the gray shaded areas show the results when the
true and the fitted event rates are computed using the same set of
migration matrices, while the solid lines show the results when a
different target is used to compute the matrices used in the true and
the fitted event rates. In our particular example, oxygen is used to
generate the matrices and cross-sections used to compute the true rates, and these are then fitted using carbon
migration matrices and cross-sections. As it can be seen from the figure, there is a
shift in the best fit for the atmospheric mass splitting from $2.45\times 10^{-3}\; \textrm{eV}^2$ to
$\sim 2.49\times 10^{-3}\; \textrm{eV}^2$. There is also a shift in
the best fit for the mixing angle, although in this case the effect is
minor. The value of the $\chi^2$ found at the best fit is also shown, together with the number of degrees of freedom in the analysis, $n-p$, where $n$ is the number of energy bins and $p$ is the number of parameters that are being determined from the fit. Finally, it is worth mentioning that we find the size of this effect to be practically negligible if the GiBUU event generator is used to produce the migration matrices and cross sections instead of GENIE. 
%
%

\subsection{Impact of different nuclear models}
\label{subsec:res-gen}
\begin{figure*}[hbtn!]
\subfigure[~No calibration error]{\includegraphics[width=1.0\columnwidth]{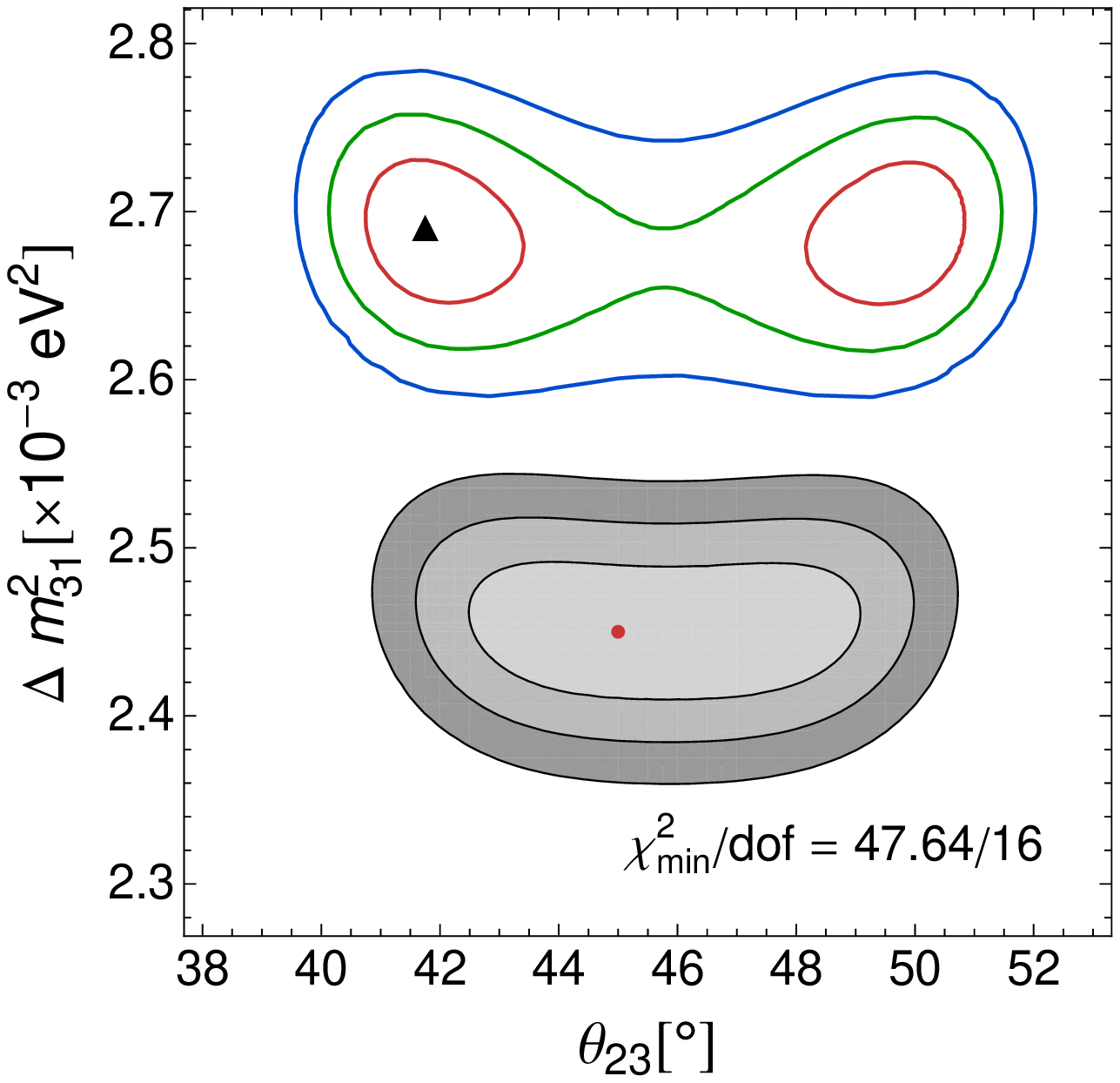}\label{fig:gen1}}
\subfigure[~5\% calibration error]{\includegraphics[width=1.0\columnwidth]{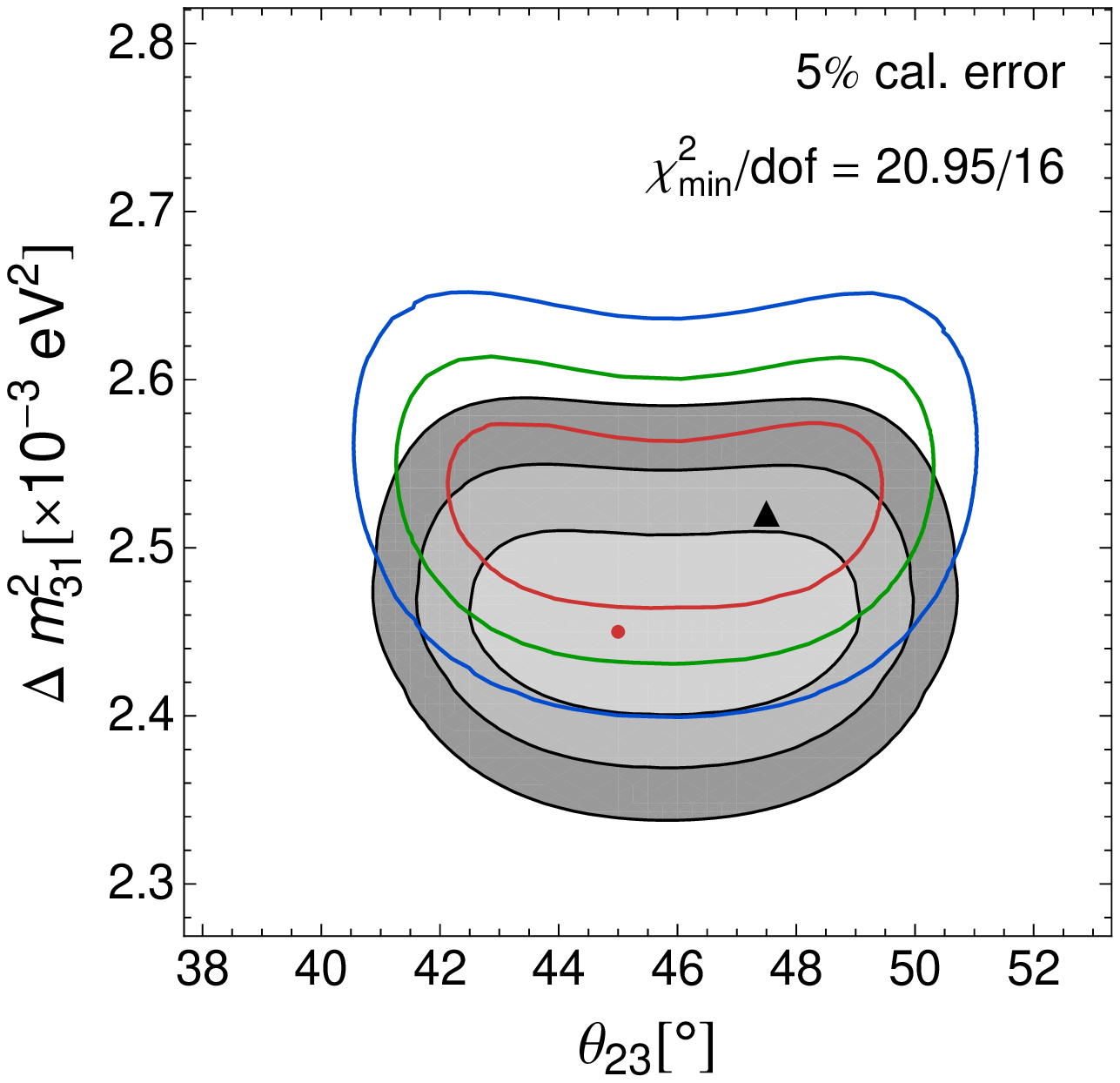}\label{fig:gen2}}
\caption{(color online)\textit{ Impact on the results if a different
    generator is used to compute the true and fitted rates in the
    analysis. The shaded areas show the confidence regions at 1, 2 and
    3$\sigma$ that would be obtained in the
    $\theta_{23}-\Delta m_{31}^2$ plane if the true and fitted rates
    are generated using the same set of migration matrices (obtained
    from GiBUU, with oxygen as the target nucleus). The colored lines
    show the same confidence regions if the true rates are generated
    using matrices produced with GiBUU, but the fitted rates are
    computed using matrices produced with GENIE. Both sets of matrices
    are generated using oxygen as the target nucleus. The red dot
    indicates the true input value, while the black triangle shows the
    location of the best fit point. The value of the $\chi^2$ at the
    best fit is also shown, together with the number of degrees of freedom. In the left panel no energy scale
    uncertainty is considered, while for the right panel an energy
    scale uncertainty of $5\%$ is assumed, see text for
    details. \label{fig:generators}}}
\end{figure*}

Let us now address the impact of using a different nuclear model. For
this purpose, we take the event rates computed using migration
matrices produced with one event generator and we try to fit them
using the matrices obtained with a different generator. The difference
in the event rates when the matrices are computed with different
generators is significant, as it can be seen from the comparison of
the solid blue and dashed red lines in Fig.~\ref{fig:rates}. In fact,
we find that the main source of the discrepancy appears in the QE
sample already, as discussed in Sec.~\ref{sec:gen}. Therefore, a large effect in the fit to
the oscillation parameters should be expected in this case. The
results are shown in Fig.~\ref{fig:generators}. Again in this case,
the shaded regions show the confidence regions when the same set of
matrices is used to compute the true and the fitted event rates. For
the solid lines, on the other hand, we compute the true event rates
using matrices produced with GiBUU, and try to fit them with the event
rates computed using the matrices from GENIE. In all cases, oxygen is
chosen as the target nucleus. We find that the difference between the
two models is so large that the best fit for the atmospheric mass
splitting takes place around $ 2.69\times 10^{-3}\;\textrm{eV}^2$, as
shown in Fig.~\ref{fig:gen1}. Such large value would be in strong
tension with the presently allowed region at $3\sigma$ from global
fits, see for instance Ref.~\cite{GonzalezGarcia:2012sz}. The fit
would disfavor the true input value for the mass splitting at much
more than $3\sigma$ and for the atmospheric mixing angle at roughly
$2\sigma$.
%
%
\par In a real experiment, this would most likely be attributed to a
systematic error that has not been correctly evaluated, or to a signal
of New Physics. We focus on the first possibility here. In particular,
since we observe that the main difference in the event rates is that
the events suffer an energy shift (see Fig.~\ref{fig:rates}, or
Fig.~\ref{fig:events_qe}), we introduce a calibration error as an
additional nuisance parameter\footnote{A real experiment if equipped
  with calibration methods will find very difficult to make such an
  adjustment. Modern calibration systems have an
  error of few \% and their results are part of the oscillation
  analysis.}. In order to do so, the event rates are effectively
recomputed as:
\begin{equation}
N[E] \rightarrow N[(1+a)\,E], \nonumber
\end{equation}
where $a$ is the calibration error. An additional pull term
$(a/\sigma_a)^2$ is added to the $\chi^2$ in Eq.~\ref{eq:chi2}, where
$\sigma_a$ is the prior uncertainty for this parameter. After a
calibration error of around a 5\% is added, the resulting best fit is
in much better agreement with the true input value, as it can be seen
from Fig.~\ref{fig:gen2}. Nevertheless, there is still a significant
shift in the best fit (black triangle) with respect to the true value
(red dot), which brings it from $2.45\times 10^{-3}\;\textrm{eV}^2$ to $2.55\times 10^{-3}\;
\textrm{eV}^2$. The best fit for the mixing angle is also shifted a
couple of degrees into the second octant. The value of the $\chi^2$ at
the minimum over the number of degrees of freedom is also computed, and found to be $ \chi^2_{min}/\rm{dof}\sim 21/16$.
We find that the largest contribution to the minimum of the $\chi^2$ comes from the tension in the energy bins below $\sim 0.5$~GeV between the two event distributions. This can be understood from the comparison of the event histograms in
Fig.~\ref{fig:rates}, where the differences between the solid
and dashed lines are observed to be largest in this energy region. In order to accommodate the large differences in these bins, the associated nuisance parameters during the $\chi^2$ minimization tend to take large values. As a consequence, their respective pull-terms will significantly contribute to the final $\chi^2$. Details on the systematics implementation can be found in App.~A.

\subsection{Impact of multinucleon contributions}
\begin{figure*}[htbn!]
\subfigure[~Results using GiBUU matrices and cross sections]{\includegraphics[width=1.0\columnwidth]{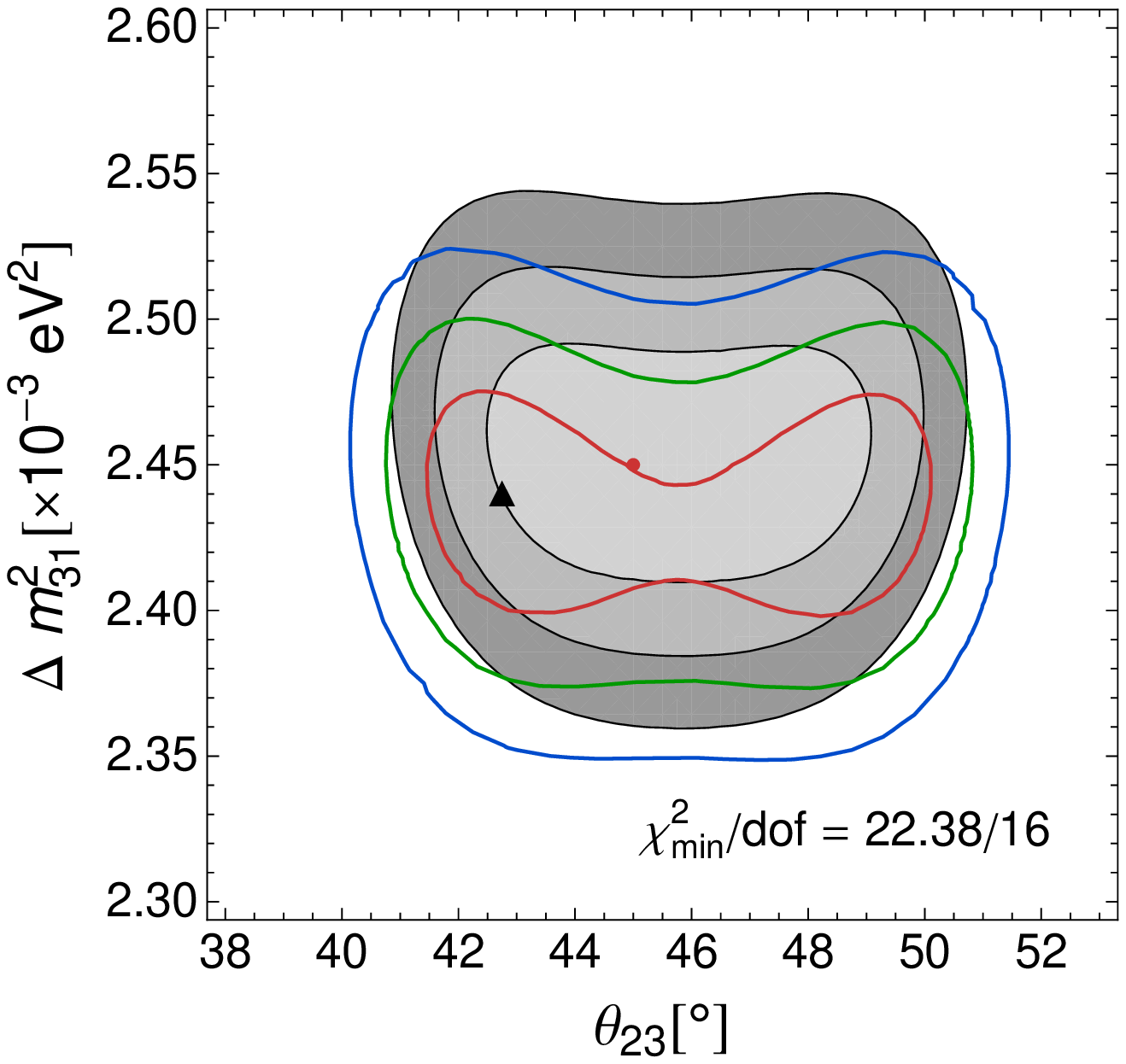}\label{fig:mec1}}
\subfigure[~Results using GENIE matrices and cross sections]{\includegraphics[width=1.0\columnwidth]{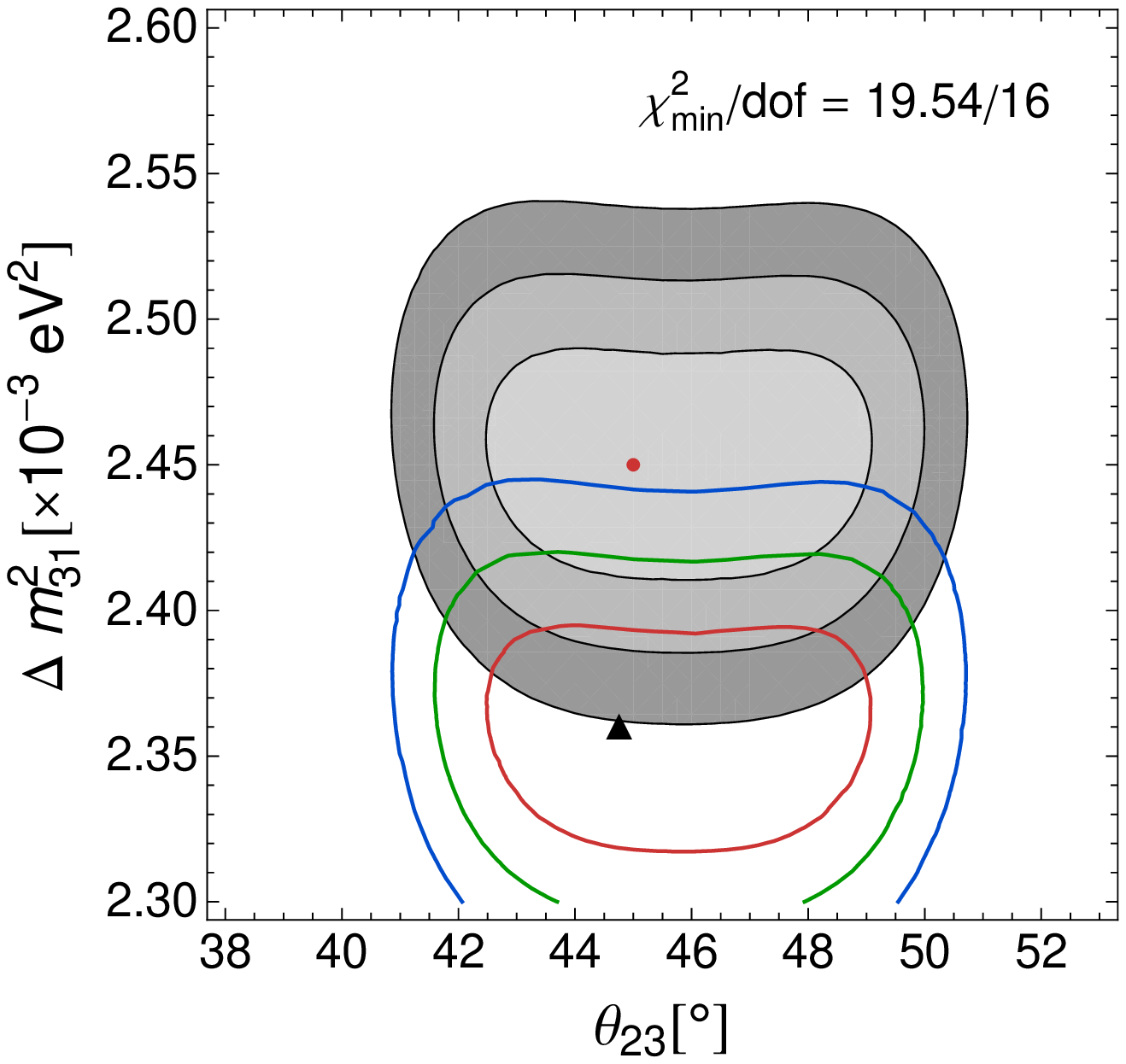}\label{fig:mec2}}
\caption{(color online)\textit{ Impact on the results if MEC/2p2h
    interactions are removed from the fit. The shaded areas include
    all contributions to QE-like events both in the true and fitted
    rates. The colored lines show the confidence regions at 1,2 and
    3$\sigma$ that would be obtained if the true rates are
    generated including MEC/2p2h interactions but they are removed from
    the fitted rates. The results are shown for the GiBUU (left panel)
    and for the GENIE (right panel) event generators. In each case,
    the red dot indicates the true input value, while the black
    triangle shows the location of the best fit point. The value of
    the $\chi^2$ at the best fit is also explicitly shown, together with the number of degrees of freedom. No energy
    calibration uncertainty has been assumed in this case for any of
    the panels.\label{fig:mec}}}
\end{figure*}

Recently, a lot of attention has been drawn to the MiniBooNE
experiment and the extraction of the value of the axial mass from the
QE data. It seems that the $\sim 35\%$ disagreement between the value of the axial mass 
extracted from the results in\footnote{Recently, the MINERvA collaboration has also observed deviations from the expected result within the RFG model approach~\cite{Fiorentini:2013ezn}.} Ref.~\cite{AguilarArevalo:2010zc} and the one obtained from previous experiments may be
explained (at least partially) if multi-nucleon contributions are included in the analysis, 
see for instance Refs.~\cite{Martini:2009uj,Nieves:2011yp,Martini:2011wp}. In this
section, we evaluate the impact that neglecting MEC/2p2h contributions
may have on the extraction of neutrino oscillation parameters. The
results are shown in Fig.~\ref{fig:mec} when the migration matrices
are computed using GiBUU (left panel) and GENIE (right panel). In both
cases, oxygen is used as the target nucleus. The shaded areas show the
confidence regions at 1,\,2 and $3\,\sigma$ CL when all contributions to
the QE-like event sample are considered as in
Eq.~\ref{eq:events-QElike}, and the same set of migration matrices is
used to compute the true and fitted event rates. For the solid lines,
on the other hand, the fit is done when the MEC/2p2h contribution is
completely removed from the fitted event rates. As expected, the
impact on the confidence regions is rather relevant, and induces large
deviation for the best fit values of the oscillation parameters,
between 1 and 3\,$\sigma$ depending on the event generator that is
used to produce the migration matrices.
%
%
\par A final comment is in order here. As it was shown previously in the
literature (see for instance Refs.~\cite{Martini:2012fa,Nieves:2012yz,Lalakulich:2012hs,Martini:2012uc,Mosel:2013fxa}) and confirmed here, the energy dependence of the QE-like
event sample is very different when MEC/2p2h contributions are
included. We have shown in this section that the impact of multi-nucleon
contributions on the extraction of oscillation parameters can be rather
large in the disappearance channels. Similar effects are in principle expected in
the appearance channels as well. This could have a significant impact on the
sensitivity to CP violation at future oscillation experiments (as mentioned, for instance, in Refs.~\cite{Lalakulich:2012hs,Martini:2012uc,Mosel:2013fxa,Nieves:2013fr}), since the
sensitivity to CP violation in neutrino oscillations comes from the
analysis of \emph{both} the energy dependence of the signal \emph{and}
the comparison between neutrino and antineutrino rates. Both GENIE and GiBUU have tuned their MEC/2p2h interactions to the
MiniBooNE data, and their results for MEC/2p2h contributions with oxygen
and carbon therefore give exactly the same results. However, \emph{a priori} there is no reason to think that these effects should be the
same for different nuclei. As for antineutrinos, there are currently very few measurements available. The MiniBooNE collaboration has recently reported some measurements in the antineutrino channel, where again it seems that MEC/2p2h may play a leading role~\cite{AguilarArevalo:2013hm}. This result has also been confirmed by the MINERvA collaboration~\cite{Fields:2013zhk}. Nevertheless, we would like to stress out the fact that the current proposals for the next generation of neutrino experiments would use either water (T2HK~\cite{Abe:2011ts} or ESS$\nu$SB~\cite{Baussan:2013zcy}, for instance) or liquid argon (LBNE~\cite{CDR,Adams:2013qkq} and LBNO~\cite{Stahl:2012exa}) detectors, for which there are practically no measurements available at the relevant neutrino energies. Again in this case, theoretical calculations show that in principle one should not expect these effects to be similar for neutrinos and antineutrinos.~\cite{Martini:2010ex,Amaro:2011aa,Nieves:2013fr,Martini:2013sha} and may be even larger for the latter, see for instance Refs.~\cite{Amaro:2011aa,Nieves:2013fr}.

\section{Conclusions}
\label{sec:concl}
%
Nuclear effects in neutrino interactions will be one of the leading
sources of systematical errors in future neutrino-beam oscillation
experiments. Already in the current T2K appearance result they are
among the largest contributors to the overall systematic error
budget~\cite{Abe:2013xua}. In this paper we try to estimate the size
of the systematic error associated with theoretical models of nuclear
effects as embodied by event generators, specifically GENIE and
GiBUU. Apart from providing a quantitative estimate we also developed a
methodological framework which lends itself to be extended to a
larger class of event generators and in principle also to CP violation
studies.
\par Given that LBNE has chosen argon as the detector material, one question is
whether changing the nuclear target will have a profound impact on the
ability to extract oscillation physics. To get a first glimpse of an
answer we study the $\nu_\mu$ disappearance channel and determine the
bias resulting from simulating data with oxygen as target and fitting
those data with a carbon interaction model. The results of this
experiment are shown in Fig.~\ref{fig:targets2} and the quantitative
findings are summarized in Tab.~\ref{tab:summary}, which correspond to
a 1$\,\sigma$ bias in $\Delta m^2_{31}$. These results are only an
indication, but it is noteworthy that most nuclear models have been
tuned on carbon data and thus, the generators can be expected to be
most accurate for nuclei around carbon. Nevertheless, it is somewhat
surprising that a small step in atomic mass from $A=12$ to $A=16$
leads already to a sizable bias. What would happen if one tried
to extrapolate to argon with $A=40$ remains pure speculation.
%
\begin{table*}
\begin{center}
\begin{tabular}{l@{\quad}l@{\quad}|@{\quad}c@{\quad}c@{\quad}c@{\quad}c@{\quad}c@{\quad}} 
True & Fitted & $\theta_{23,min} $ & $\Delta
m^2_{31,min}[\textrm{eV}^2] $ & $\chi^2_{min}$ & $\sigma_a$ & Fig.~no. \\ \hline GENIE ($^{16}$O) & GENIE ($^{12}$C) & 
$44^\circ$ & 2.49$\times 10^{-3}$ & 2.28 & -- & \ref{fig:targets} \\ \hline
\multirow{2}{*}{GiBUU ($^{16}$O)} & \multirow{2}{*}{GENIE ($^{16}$O)}
   & $41.75^\circ$ & 2.69$\times 10^{-3}$ & 47.64 & --  & \ref{fig:gen1}
\cr & & $47^\circ$ & 2.55$\times 10^{-3}$ & 20.95 & 5\% &
\ref{fig:gen2} \\ \hline GiBUU ($^{16}$O) & GiBUU ($^{16}$O) w/o MEC &
$42.5^\circ$ & 2.44$\times 10^{-3}$ & 22.38 & -- & \ref{fig:mec1}
\\ \hline GENIE ($^{16}$O) & GENIE ($^{16}$O) w/o MEC & 
$44.5^\circ$ & 2.36$\times 10^{-3}$ & 19.54 & -- & \ref{fig:mec2} \\ \hline
\end{tabular}
\caption{\textit{Summary of the main impact on the oscillation parameters for
  the different scenarios studied in this work. The true values for
  the disappearance oscillation parameters are $\theta_{23}=45^\circ$ and $\Delta
  m^2_{31}=2.45\times10^{-3}\,\mathrm{eV}^2$. The number of degrees of
  freedom in the fit is $n-p=16$, where $n$ is the number of energy bins and $p$ is the number of oscillation parameters that are being estimated from the fit. Here, $\sigma_a$ represents the prior uncertainty assumed for an energy calibration error, whose implementation is described in Sec.~\ref{subsec:res-gen}.} \label{tab:summary}}
\end{center}
\end{table*}
%
\par Interestingly, we find even for carbon very large differences in the
shape of the QE event rate spectrum between GENIE and GiBUU which we
trace back to differences in the implementation of final state
interactions, see Sec.~\ref{sec:gen}. These differences are
large enough to introduce a bias in the mass splitting of many
standard deviations. The resulting minimum of the $\chi^2$ would also be very large as
shown in Fig.~\ref{fig:gen1}. Introducing an uncertainty on the energy
scale of 5\% reduces the resulting tension and brings also the
$\chi^2$ back to acceptable levels, but still leaves a 1\,$\sigma$
bias in both the mass splitting and mixing angle, see
Fig.~\ref{fig:gen2}. Besides, it is also not clear whether an oscillation experiment would 
have any freedom left to include such a large calibration error in the fit.
\par Finally, the recent MiniBooNE results seem to imply that multi-nucleon effects
play an important role for neutrino energies $E_\nu\sim \mathcal{O}(\mathrm{GeV})$. There is a large
variety of models in the literature trying to describe these effects. Therefore, we test
the effect of removing the multi-nucleon correlation in the fit of
data which has been generated including those. This is again a case
where the two generators produce different effects -- in GiBUU the
mixing angle is most affected whereas in GENIE it is the mass
splitting seeing the bulk of the effect, which is obvious from
Fig.~\ref{fig:mec}.
\par In summary, we find that changing nuclear targets, the used event
generator or the implementation of multi-nucleon effects, each leave a
bias comparable to the statistical errors in the determination of the
mixing parameters, as illustrated in Tab.~\ref{tab:summary}. Any
experiment aiming at high precision measurements of oscillation
parameters like the leptonic CP phase will have to develop a strategy
to deal with these uncertainties in a transparent fashion. One
important step in this direction would be to make the event generators 
accessible to the community. We have only considered
light targets like carbon and oxygen and it is unclear how to
extrapolate to heavier targets like argon without additional data. The
methods presented here are well suited to be extended to experiments
aiming to determine the CP phase and the neutrino mass hierarchy.

\section*{Acknowledgements}

The authors would like to thank S.~Dytman, U.~Mosel and O.~Lalakulich for their support, which lead to successful simulation results. We would also like to thank D.~Meloni, J.~Nieves, N. Rocco and M.~Vicente Vacas for useful discussions, and O.~Benhar for the precious discussions and for careful reading of this manuscript. This work has been supported by the U.S. Department of Energy under award number \protect{DE-SC0003915}. 

\clearpage
\appendix
\section{$\chi^2$ implementation}
\label{app:chi2}
%
In this appendix we briefly describe the details on the implementation of the $\chi^2$ and the inclusion of systematic uncertainties. 
All fits to the oscillation parameters presented in Sec.~\ref{sec:results} are performed using GLoBES~\cite{Huber:2004ka,Huber:2007ji}. 
Unless otherwise stated, the true values of the oscillation parameters are set to the following:
\begin{eqnarray}
\nonumber \theta_{12} = & 33.2^\circ ~~ & \Delta m^2_{21}  = 7.64\times 10^{-5} \, \textrm{eV}^2 \\
\nonumber \theta_{13} = & 9^\circ ~~ &  \Delta m^2_{31}  = 2.45\times 10^{-3} \, \textrm{eV}^2 \\
\theta_{23} = & 45^\circ ~~ &  \delta  =  0^\circ \,.
\label{eq:oscparams}
\end{eqnarray}
A $\chi^2$ analysis is done to extract the best fit values for the oscillation parameters as well as the allowed confidence regions at 1, 2 and 3$\sigma$ CL. For each energy bin $i$ and detector $D$, a contribution to the $\chi^2$ is computed as:
\begin{align}
\nonumber &\chi^2_{i,D} =  \\
\nonumber &2\bigg(T_{i,D}(\theta,\xi)-O_{i,D}+O_{i,D} \ln \frac{O_{i,D}}{T_{i,D}(\theta,\xi))} \bigg) \,,\\
\end{align}
with
\[T_{i,D}(\theta,\xi)= \left(1+\xi_n + \xi_{\phi,i}\right)N_{i,D}(\theta).\]
Here, $O_{i,D}$ ($T_{i,D}$) refer to the true (fitted) event rates observed at a detector $D$ in an energy bin $i$, $\theta$ indicates the dependence on the test values for the oscillation parameters, and $ \xi_{\phi,i}$ and $\xi_n $ stand for the nuisance parameters associated to flux and normalization uncertainties, respectively. It should be noted that $O_{i,D}$ depends only on the true values assumed for the oscillation parameters, while $T_{i,D}$ depends on the pair of values we are testing as well as on the nuisance parameters. In addition, the nuisance parameter associated to the normalization error is bin-to-bin correlated, while the one associated to the flux uncertainty is bin-to-bin uncorrelated. These will help to accommodate the normalization and shape differences in the event rates due to different nuclear models. The final $\chi^2$ reads:
%
\begin{align}
\nonumber &\chi^2 =\\
\nonumber &min_{\xi}\left\{\sum_{D,i}\chi^2_{i,D}(\theta;\xi)+\left(\frac{\xi_{\phi,i}}{\sigma_\phi}\right)^2+\left(\frac{\xi_{n}}{\sigma_{n}}\right)^2\right\},\\
\label{eq:chi2}
\end{align}
%
where the two last terms are the pull-terms associated to the nuisance parameters $\xi$, and $\sigma_\xi$ is the prior uncertainty assumed for each systematic error (which have been set to 20\% in all cases). All oscillation parameters are kept fixed during the fit. Since the analysis is done for the disappearance $\nu_\mu \rightarrow \nu_\mu$ channel only, we believe that marginalization will not have a relevant impact on our results. Finally, the sum in Eq.~\ref{eq:chi2} is done over 100~MeV bins between 0.2 and 2.0 GeV. Nevertheless, the migration matrices and cross-sections are computed up to $\sim5$ GeV in true neutrino energy in order to get the full contribution from the high-energy tail of the flux.

%
\section{Migration matrices and event distributions}
\label{app:matrices}

\begin{figure*}[htbn!]
\subfigure[~Matrix for QE events]{\includegraphics[width=1.0\columnwidth]{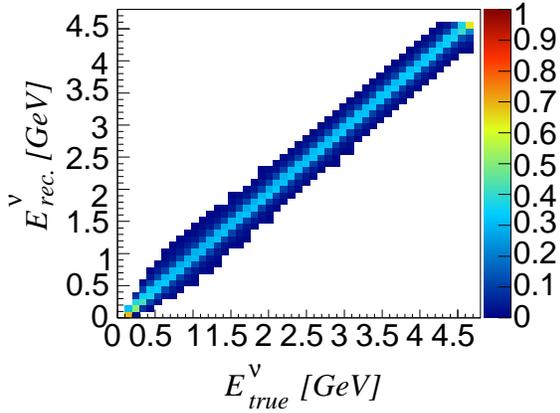}\label{fig:genie_matrix_qe}}
\subfigure[~Matrix for RES events]{\includegraphics[width=1.0\columnwidth]{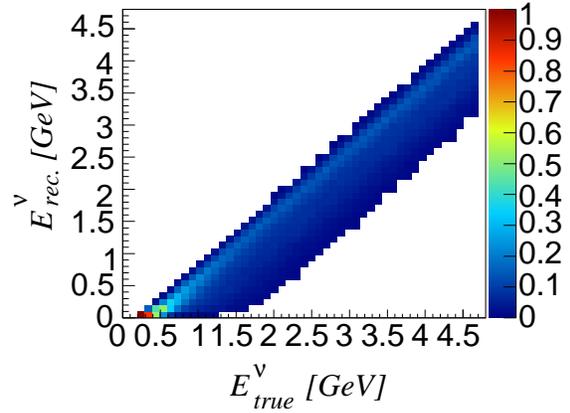}\label{fig:genie_matrix_res}}
\subfigure[~Matrix for MEC/2p2h events]{\includegraphics[width=1.0\columnwidth]{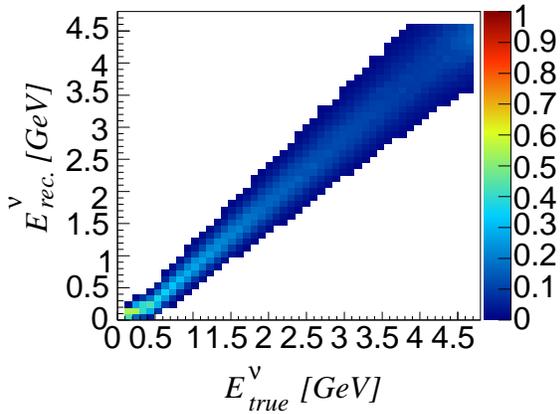}\label{fig:genie_matrix_2p2h}}
\subfigure[~Matrix for non-RES events]{\includegraphics[width=1.0\columnwidth]{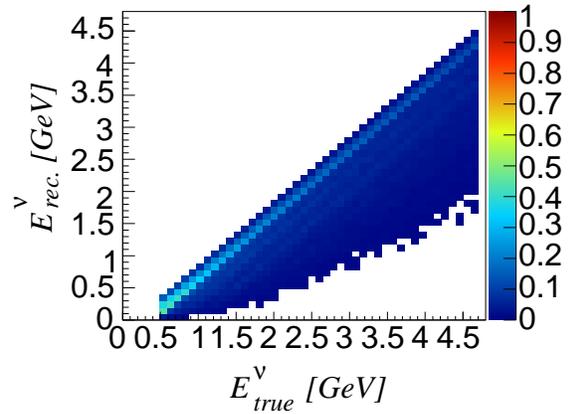}\label{fig:genie_matrix_coh}}
\caption{(color online) \textit{Two-dimensional migration matrices ($M_{ij}$), for QE(a), RES(b), MEC/2p2h(c), non-RES(d) for GENIE, using $^{16}$O as the target nucleus. }}
\label{fig:genie_matrix}
\end{figure*}

\begin{figure*}[htbn!]
\subfigure[~Matrix for QE events]{\includegraphics[width=1.0\columnwidth]{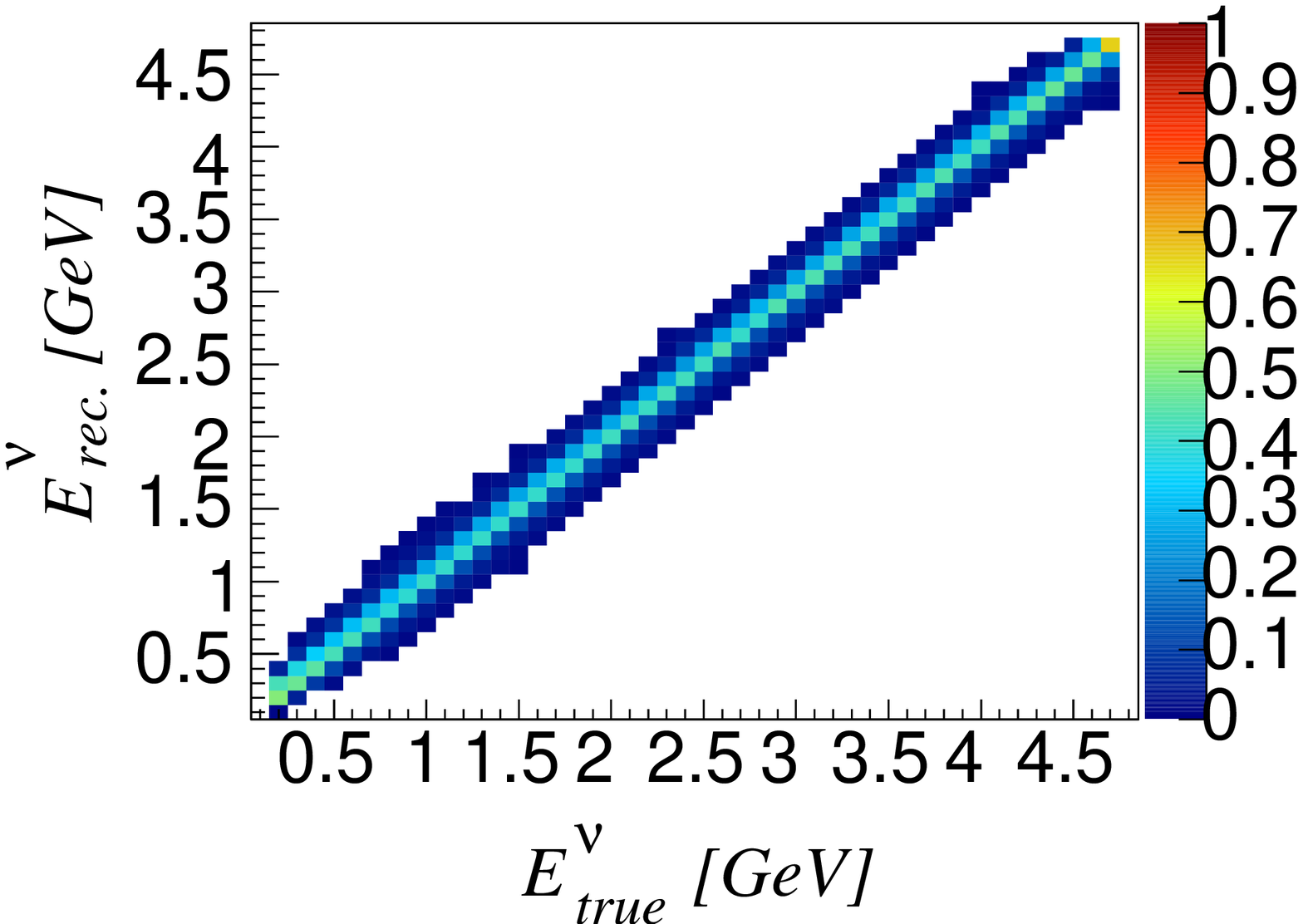}\label{fig:gibuu_matrix_qe}}
\subfigure[~Matrix for RES events]{\includegraphics[width=1.0\columnwidth]{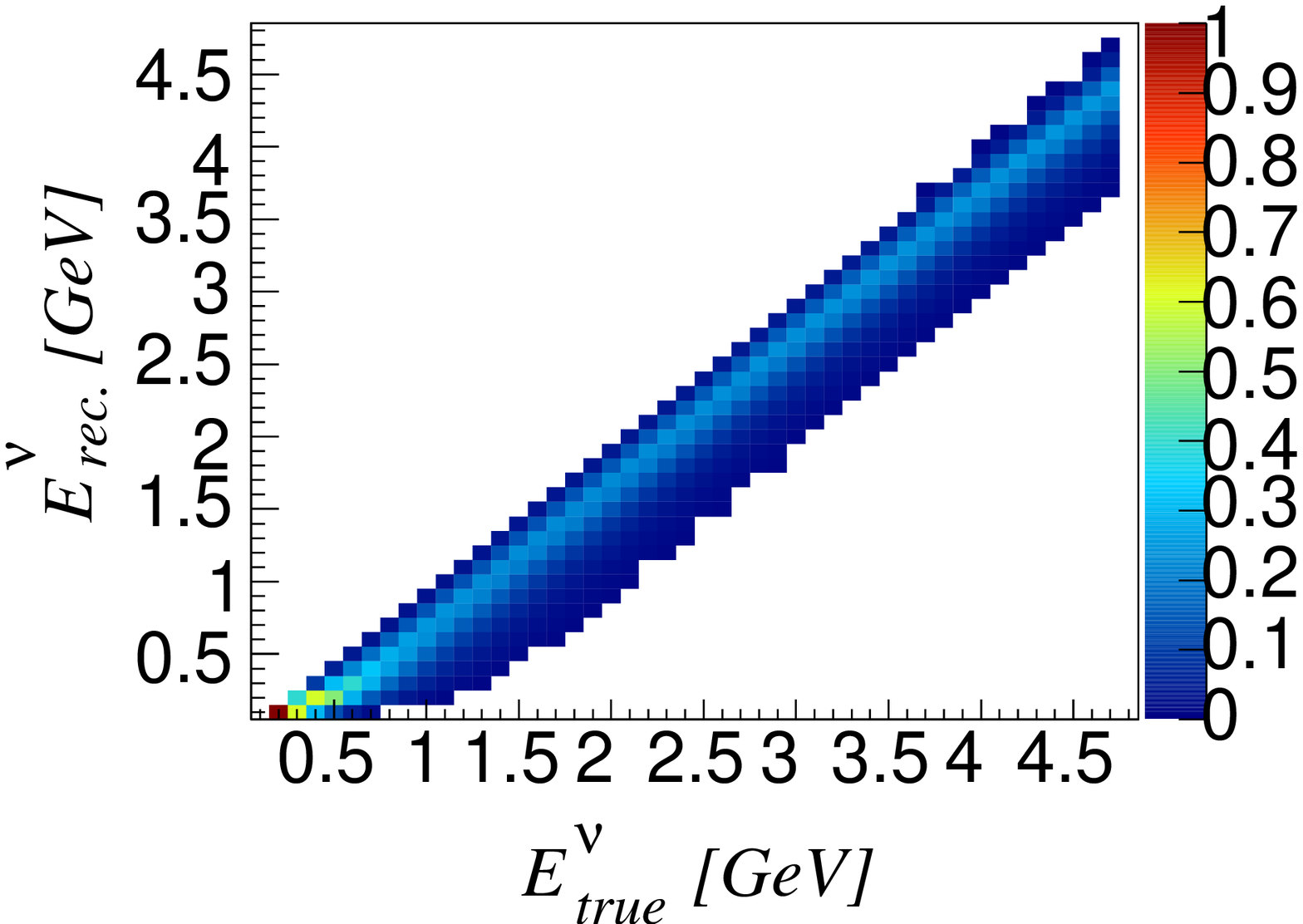}\label{fig:gibuu_matrix_res}}
\subfigure[~Matrix for MEC/2p2h events]{\includegraphics[width=1.0\columnwidth]{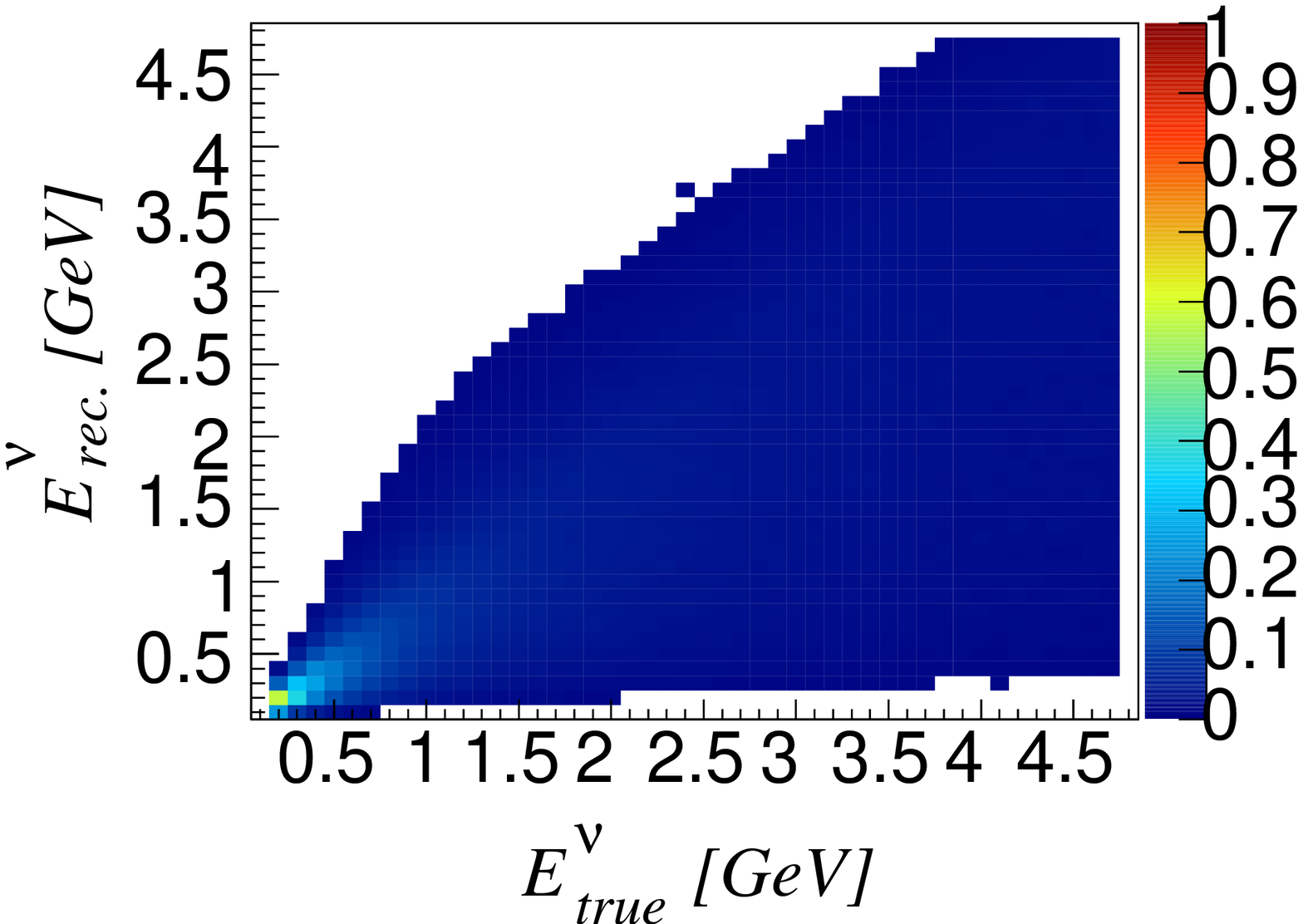}\label{fig:gibuu_matrix_2p2h}}
\subfigure[~Matrix for non-RES events]{\includegraphics[width=1.0\columnwidth]{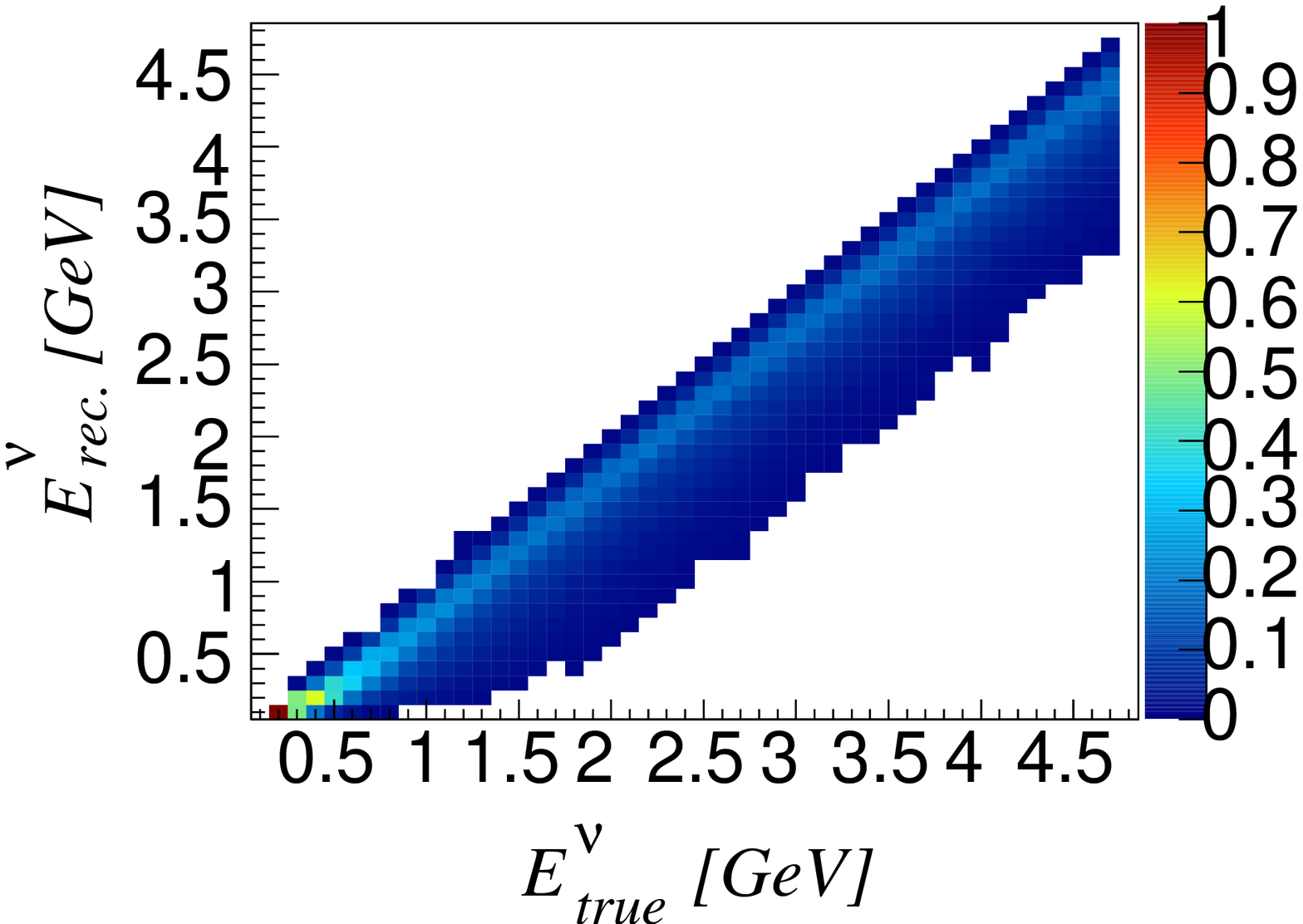}\label{fig:gibuu_matrix_coh}}
\caption{(color online) \textit{Two-dimensional migration matrices ($M_{ij}$), for QE(a), RES(b), MEC/2p2h(c) and non-RES(d) for GiBUU, using $^{16}$O as the target nucleus. }}
\label{fig:gibuu_matrix}
\end{figure*}

\begin{figure*}[htbn!]
\begin{center}
\subfigure[~RES events]{\includegraphics[width=1.0\columnwidth]{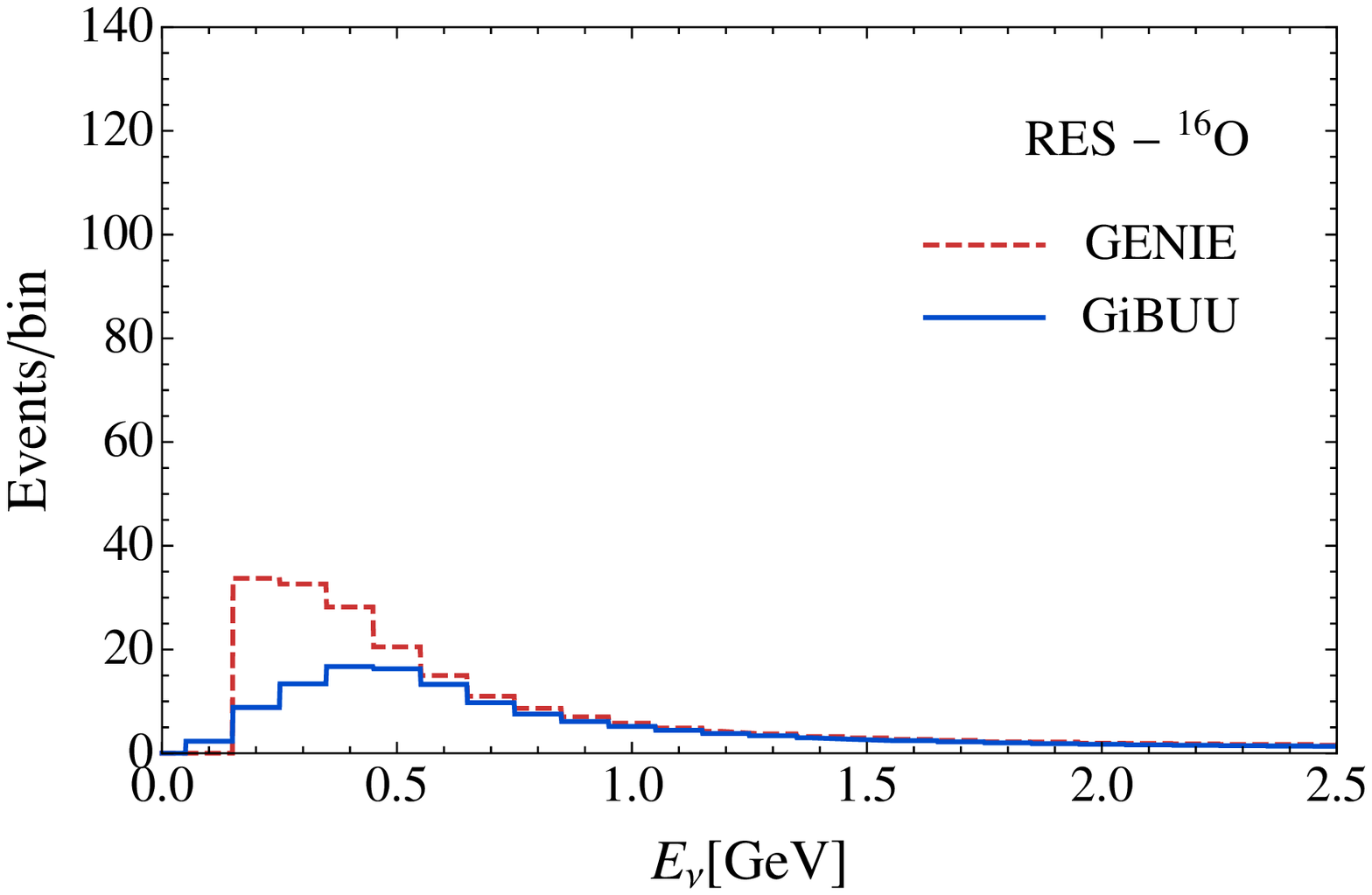}\label{fig:events_res}}
\subfigure[~MEC/2p2h events]{\includegraphics[width=1.0\columnwidth]{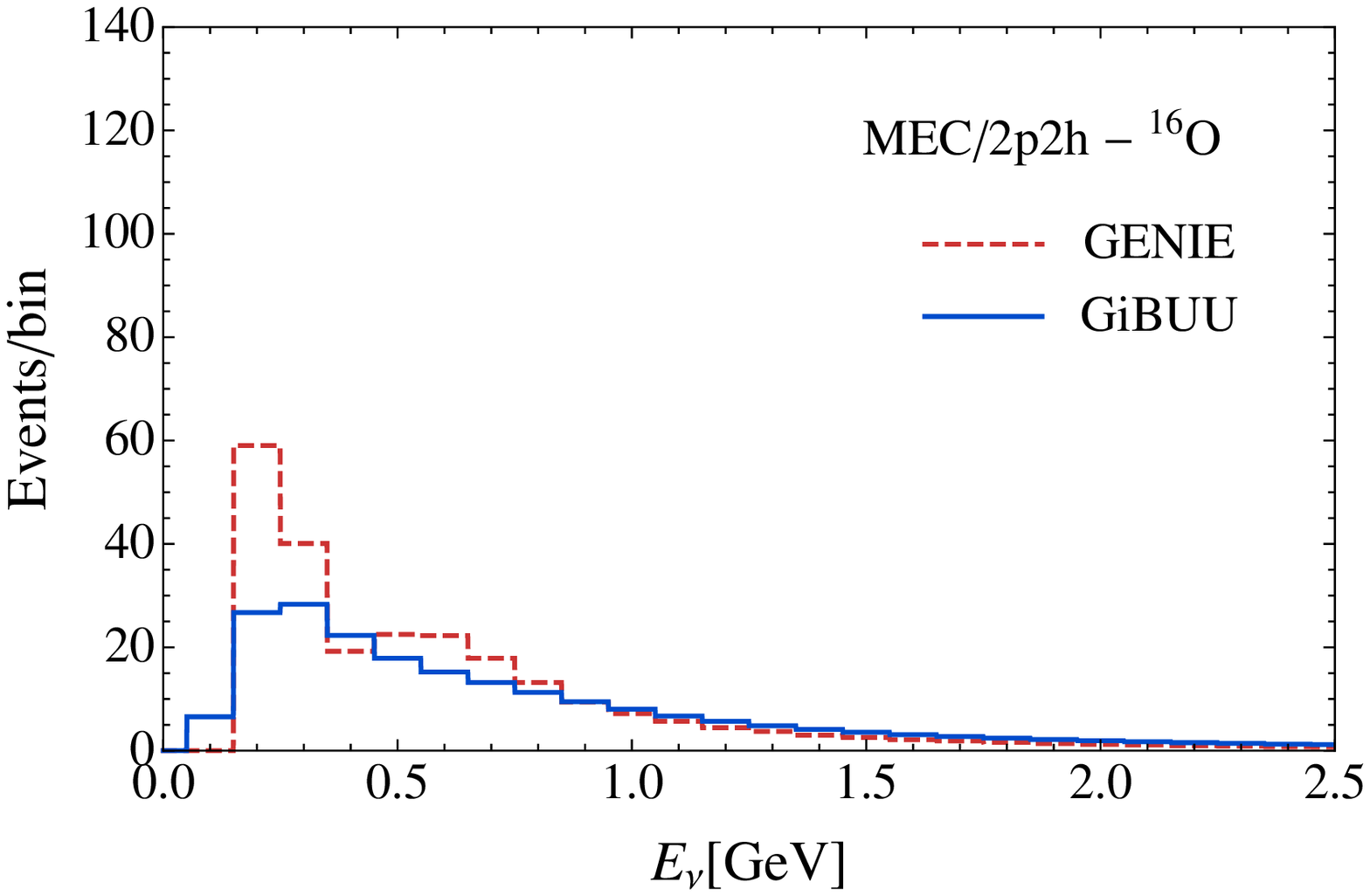}\label{fig:events_2p2h}}
\subfigure[~non-RES events]{\includegraphics[width=1.0\columnwidth]{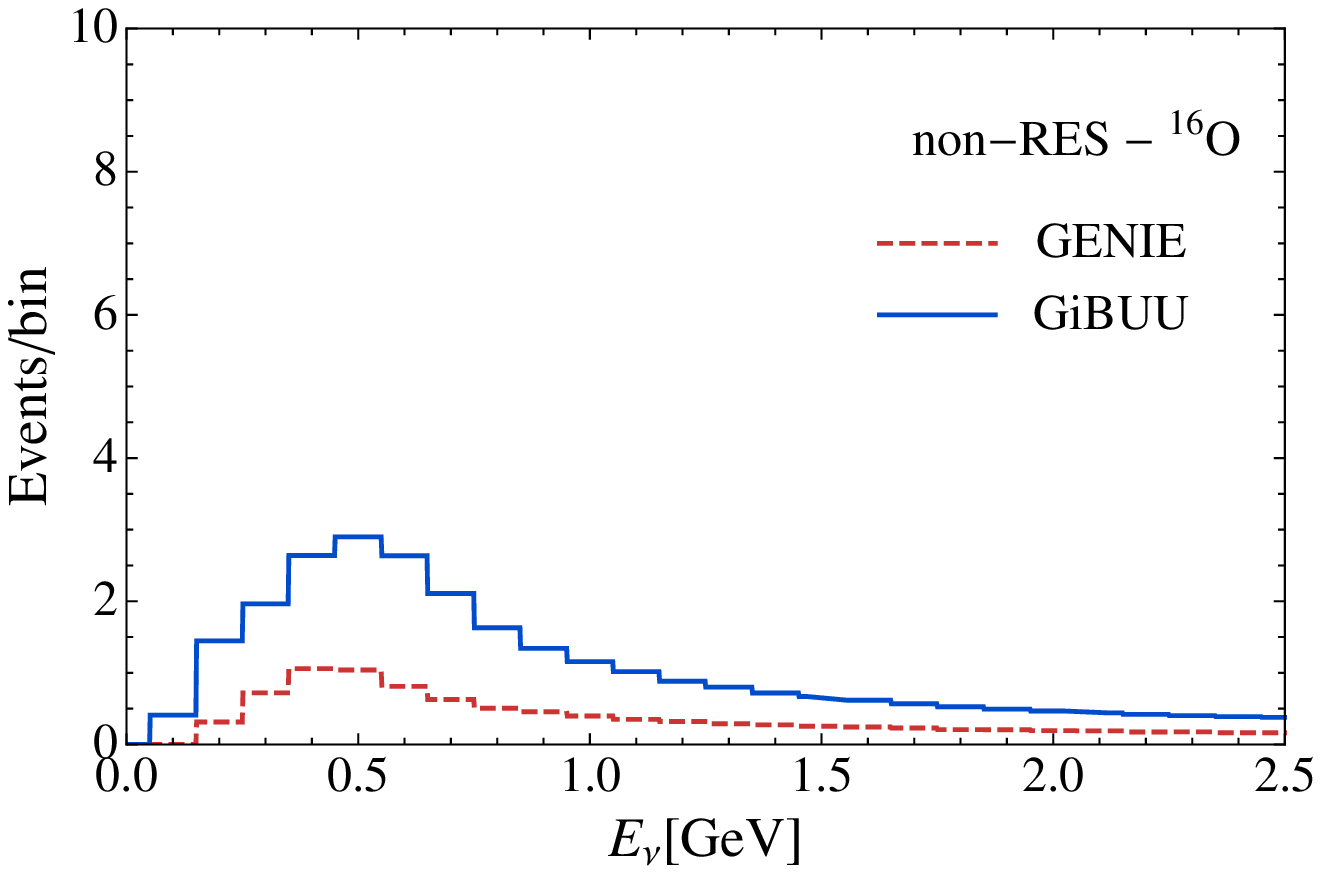}\label{fig:events_coh}}
\caption{(color online) \textit{Event distributions as a function of the reconstructed neutrino energy for both GENIE (dashed red lines) and GiBUU (solid blue lines), for the non-QE contributions to the QE-like sample. In both cases, oxygen is used as the target nucleus to obtain the cross sections and migration matrices. The different panels show the event distributions for RES(a), MEC/2p2h(b) and non-RES(c) events. The oscillation parameters have been set to their values in Eq.~\ref{eq:oscparams}, and detection efficiencies have already been accounted for. }}
\label{fig:event_distribution_carbon}
\end{center}
\end{figure*}

In this appendix we show the complete set of migration matrices used in the oscillation analysis for GENIE (Fig.~\ref{fig:genie_matrix}) and GiBUU (Fig.~\ref{fig:gibuu_matrix}) using oxygen as the target nucleus. The number of events as function of reconstructed neutrino energy for the various neutrino interaction modes, as described in Sec.~\ref{sec:gen}, is also shown in Fig.~\ref{fig:event_distribution_carbon} for oxygen. The results for carbon are similar and therefore will not be shown here.

\clearpage 


\begin{thebibliography}{80}
\expandafter\ifx\csname natexlab\endcsname\relax\def\natexlab#1{#1}\fi
\expandafter\ifx\csname bibnamefont\endcsname\relax
  \def\bibnamefont#1{#1}\fi
\expandafter\ifx\csname bibfnamefont\endcsname\relax
  \def\bibfnamefont#1{#1}\fi
\expandafter\ifx\csname citenamefont\endcsname\relax
  \def\citenamefont#1{#1}\fi
\expandafter\ifx\csname url\endcsname\relax
  \def\url#1{\texttt{#1}}\fi
\expandafter\ifx\csname urlprefix\endcsname\relax\def\urlprefix{URL }\fi
\providecommand{\bibinfo}[2]{#2}
\providecommand{\eprint}[2][]{\url{#2}}

\bibitem[{\citenamefont{Abe et~al.}(2011{\natexlab{a}})}]{Abe:2011sj}
\bibinfo{author}{\bibfnamefont{K.}~\bibnamefont{Abe}} \bibnamefont{et~al.}
  (\bibinfo{collaboration}{T2K Collaboration}),
  \bibinfo{journal}{Phys.Rev.Lett.} \textbf{\bibinfo{volume}{107}},
  \bibinfo{pages}{041801} (\bibinfo{year}{2011}{\natexlab{a}}),
  \eprint{1106.2822}.

\bibitem[{\citenamefont{Abe et~al.}(2012)}]{Abe:2011fz}
\bibinfo{author}{\bibfnamefont{Y.}~\bibnamefont{Abe}} \bibnamefont{et~al.}
  (\bibinfo{collaboration}{DOUBLE-CHOOZ Collaboration}),
  \bibinfo{journal}{Phys.Rev.Lett.} \textbf{\bibinfo{volume}{108}},
  \bibinfo{pages}{131801} (\bibinfo{year}{2012}), \eprint{1112.6353}.

\bibitem[{\citenamefont{An et~al.}(2012)}]{An:2012eh}
\bibinfo{author}{\bibfnamefont{F.}~\bibnamefont{An}} \bibnamefont{et~al.}
  (\bibinfo{collaboration}{DAYA-BAY Collaboration}),
  \bibinfo{journal}{Phys.Rev.Lett.} \textbf{\bibinfo{volume}{108}},
  \bibinfo{pages}{171803} (\bibinfo{year}{2012}), \eprint{1203.1669}.

\bibitem[{\citenamefont{Ahn et~al.}(2012)}]{Ahn:2012nd}
\bibinfo{author}{\bibfnamefont{J.}~\bibnamefont{Ahn}} \bibnamefont{et~al.}
  (\bibinfo{collaboration}{RENO collaboration}),
  \bibinfo{journal}{Phys.Rev.Lett.} \textbf{\bibinfo{volume}{108}},
  \bibinfo{pages}{191802} (\bibinfo{year}{2012}), \eprint{1204.0626}.

\bibitem[{\citenamefont{Itow}(2002)}]{Itow:2002rk}
\bibinfo{author}{\bibfnamefont{Y.}~\bibnamefont{Itow}},
  \bibinfo{journal}{Nucl.Phys.Proc.Suppl.} \textbf{\bibinfo{volume}{112}},
  \bibinfo{pages}{3} (\bibinfo{year}{2002}).

\bibitem[{\citenamefont{Harris et~al.}(2004)}]{Harris:2004iq}
\bibinfo{author}{\bibfnamefont{D.}~\bibnamefont{Harris}} \bibnamefont{et~al.}
  (\bibinfo{collaboration}{MINERvA Collaboration}) (\bibinfo{year}{2004}),
  \eprint{hep-ex/0410005}.

\bibitem[{\citenamefont{Huber et~al.}(2008)\citenamefont{Huber, Mezzetto, and
  Schwetz}}]{Huber:2007em}
\bibinfo{author}{\bibfnamefont{P.}~\bibnamefont{Huber}},
  \bibinfo{author}{\bibfnamefont{M.}~\bibnamefont{Mezzetto}}, \bibnamefont{and}
  \bibinfo{author}{\bibfnamefont{T.}~\bibnamefont{Schwetz}},
  \bibinfo{journal}{JHEP} \textbf{\bibinfo{volume}{0803}}, \bibinfo{pages}{021}
  (\bibinfo{year}{2008}), \eprint{0711.2950}.

\bibitem[{\citenamefont{Fernandez-Martinez and
  Meloni}(2011)}]{FernandezMartinez:2010dm}
\bibinfo{author}{\bibfnamefont{E.}~\bibnamefont{Fernandez-Martinez}}
  \bibnamefont{and} \bibinfo{author}{\bibfnamefont{D.}~\bibnamefont{Meloni}},
  \bibinfo{journal}{Phys.Lett.} \textbf{\bibinfo{volume}{B697}},
  \bibinfo{pages}{477} (\bibinfo{year}{2011}), \eprint{1010.2329}.

\bibitem[{\citenamefont{Meloni and Martini}(2012)}]{Meloni:2012fq}
\bibinfo{author}{\bibfnamefont{D.}~\bibnamefont{Meloni}} \bibnamefont{and}
  \bibinfo{author}{\bibfnamefont{M.}~\bibnamefont{Martini}},
  \bibinfo{journal}{Phys.Lett.} \textbf{\bibinfo{volume}{B716}},
  \bibinfo{pages}{186} (\bibinfo{year}{2012}), \eprint{1203.3335}.

\bibitem[{\citenamefont{Coloma et~al.}(2013)\citenamefont{Coloma, Huber, Kopp,
  and Winter}}]{Coloma:2012ji}
\bibinfo{author}{\bibfnamefont{P.}~\bibnamefont{Coloma}},
  \bibinfo{author}{\bibfnamefont{P.}~\bibnamefont{Huber}},
  \bibinfo{author}{\bibfnamefont{J.}~\bibnamefont{Kopp}}, \bibnamefont{and}
  \bibinfo{author}{\bibfnamefont{W.}~\bibnamefont{Winter}},
  \bibinfo{journal}{Phys.Rev.} \textbf{\bibinfo{volume}{D87}},
  \bibinfo{pages}{033004} (\bibinfo{year}{2013}), \eprint{1209.5973}.

\bibitem[{\citenamefont{Martini et~al.}(2012)\citenamefont{Martini, Ericson,
  and Chanfray}}]{Martini:2012fa}
\bibinfo{author}{\bibfnamefont{M.}~\bibnamefont{Martini}},
  \bibinfo{author}{\bibfnamefont{M.}~\bibnamefont{Ericson}}, \bibnamefont{and}
  \bibinfo{author}{\bibfnamefont{G.}~\bibnamefont{Chanfray}},
  \bibinfo{journal}{Phys.Rev.} \textbf{\bibinfo{volume}{D85}},
  \bibinfo{pages}{093012} (\bibinfo{year}{2012}), \eprint{1202.4745}.

\bibitem[{\citenamefont{Nieves et~al.}(2012{\natexlab{a}})\citenamefont{Nieves,
  Sanchez, Ruiz~Simo, and Vicente~Vacas}}]{Nieves:2012yz}
\bibinfo{author}{\bibfnamefont{J.}~\bibnamefont{Nieves}},
  \bibinfo{author}{\bibfnamefont{F.}~\bibnamefont{Sanchez}},
  \bibinfo{author}{\bibfnamefont{I.}~\bibnamefont{Ruiz~Simo}},
  \bibnamefont{and}
  \bibinfo{author}{\bibfnamefont{M.}~\bibnamefont{Vicente~Vacas}},
  \bibinfo{journal}{Phys.Rev.} \textbf{\bibinfo{volume}{D85}},
  \bibinfo{pages}{113008} (\bibinfo{year}{2012}{\natexlab{a}}),
  \eprint{1204.5404}.

\bibitem[{\citenamefont{Lalakulich
  et~al.}(2012{\natexlab{a}})\citenamefont{Lalakulich, Mosel, and
  Gallmeister}}]{Lalakulich:2012hs}
\bibinfo{author}{\bibfnamefont{O.}~\bibnamefont{Lalakulich}},
  \bibinfo{author}{\bibfnamefont{U.}~\bibnamefont{Mosel}}, \bibnamefont{and}
  \bibinfo{author}{\bibfnamefont{K.}~\bibnamefont{Gallmeister}},
  \bibinfo{journal}{Phys.Rev.} \textbf{\bibinfo{volume}{C86}},
  \bibinfo{pages}{054606} (\bibinfo{year}{2012}{\natexlab{a}}),
  \eprint{1208.3678}.

\bibitem[{\citenamefont{Martini et~al.}(2013)\citenamefont{Martini, Ericson,
  and Chanfray}}]{Martini:2012uc}
\bibinfo{author}{\bibfnamefont{M.}~\bibnamefont{Martini}},
  \bibinfo{author}{\bibfnamefont{M.}~\bibnamefont{Ericson}}, \bibnamefont{and}
  \bibinfo{author}{\bibfnamefont{G.}~\bibnamefont{Chanfray}},
  \bibinfo{journal}{Phys.Rev.} \textbf{\bibinfo{volume}{D87}},
  \bibinfo{pages}{013009} (\bibinfo{year}{2013}), \eprint{1211.1523}.

\bibitem[{\citenamefont{Mosel et~al.}(2013)\citenamefont{Mosel, Lalakulich, and
  Gallmeister}}]{Mosel:2013fxa}
\bibinfo{author}{\bibfnamefont{U.}~\bibnamefont{Mosel}},
  \bibinfo{author}{\bibfnamefont{O.}~\bibnamefont{Lalakulich}},
  \bibnamefont{and}
  \bibinfo{author}{\bibfnamefont{K.}~\bibnamefont{Gallmeister}}
  (\bibinfo{year}{2013}), \eprint{1311.7288}.

\bibitem[{\citenamefont{Coloma and Huber}(2013)}]{Coloma:2013rqa}
\bibinfo{author}{\bibfnamefont{P.}~\bibnamefont{Coloma}} \bibnamefont{and}
  \bibinfo{author}{\bibfnamefont{P.}~\bibnamefont{Huber}},
  \bibinfo{journal}{Phys. Rev. Lett.} \textbf{\bibinfo{volume}{111}},
  \bibinfo{pages}{221802} (\bibinfo{year}{2013}), \eprint{1311.4506}.

\bibitem[{\citenamefont{Andreopoulos et~al.}(2010)\citenamefont{Andreopoulos,
  Bell, Bhattacharya, Cavanna, Dobson et~al.}}]{Andreopoulos:2009rq}
\bibinfo{author}{\bibfnamefont{C.}~\bibnamefont{Andreopoulos}},
  \bibinfo{author}{\bibfnamefont{A.}~\bibnamefont{Bell}},
  \bibinfo{author}{\bibfnamefont{D.}~\bibnamefont{Bhattacharya}},
  \bibinfo{author}{\bibfnamefont{F.}~\bibnamefont{Cavanna}},
  \bibinfo{author}{\bibfnamefont{J.}~\bibnamefont{Dobson}},
  \bibnamefont{et~al.}, \bibinfo{journal}{Nucl.Instrum.Meth.}
  \textbf{\bibinfo{volume}{A614}}, \bibinfo{pages}{87} (\bibinfo{year}{2010}),
  \eprint{0905.2517}.

\bibitem[{\citenamefont{Leitner
  et~al.}(2009{\natexlab{a}})\citenamefont{Leitner, Buss, and
  Mosel}}]{Leitner:2009ke}
\bibinfo{author}{\bibfnamefont{T.}~\bibnamefont{Leitner}},
  \bibinfo{author}{\bibfnamefont{O.}~\bibnamefont{Buss}}, \bibnamefont{and}
  \bibinfo{author}{\bibfnamefont{U.}~\bibnamefont{Mosel}},
  \bibinfo{journal}{Acta Phys.Polon.} \textbf{\bibinfo{volume}{B40}},
  \bibinfo{pages}{2585} (\bibinfo{year}{2009}{\natexlab{a}}),
  \eprint{0905.1644}.

\bibitem[{\citenamefont{Buss et~al.}(2012)\citenamefont{Buss, Gaitanos,
  Gallmeister, van Hees, Kaskulov et~al.}}]{Buss:2011mx}
\bibinfo{author}{\bibfnamefont{O.}~\bibnamefont{Buss}},
  \bibinfo{author}{\bibfnamefont{T.}~\bibnamefont{Gaitanos}},
  \bibinfo{author}{\bibfnamefont{K.}~\bibnamefont{Gallmeister}},
  \bibinfo{author}{\bibfnamefont{H.}~\bibnamefont{van Hees}},
  \bibinfo{author}{\bibfnamefont{M.}~\bibnamefont{Kaskulov}},
  \bibnamefont{et~al.}, \bibinfo{journal}{Phys.Rept.}
  \textbf{\bibinfo{volume}{512}}, \bibinfo{pages}{1} (\bibinfo{year}{2012}),
  \eprint{1106.1344}.

\bibitem[{\citenamefont{Delorme and Ericson}(1985)}]{Delorme:1985ps}
\bibinfo{author}{\bibfnamefont{J.}~\bibnamefont{Delorme}} \bibnamefont{and}
  \bibinfo{author}{\bibfnamefont{M.}~\bibnamefont{Ericson}},
  \bibinfo{journal}{Phys.Lett.} \textbf{\bibinfo{volume}{B156}},
  \bibinfo{pages}{263} (\bibinfo{year}{1985}).

\bibitem[{\citenamefont{Marteau et~al.}(2000)\citenamefont{Marteau, Delorme,
  and Ericson}}]{Marteau:1999jp}
\bibinfo{author}{\bibfnamefont{J.}~\bibnamefont{Marteau}},
  \bibinfo{author}{\bibfnamefont{J.}~\bibnamefont{Delorme}}, \bibnamefont{and}
  \bibinfo{author}{\bibfnamefont{M.}~\bibnamefont{Ericson}},
  \bibinfo{journal}{Nucl.Instrum.Meth.} \textbf{\bibinfo{volume}{A451}},
  \bibinfo{pages}{76} (\bibinfo{year}{2000}).

\bibitem[{\citenamefont{Martini et~al.}(2009)\citenamefont{Martini, Ericson,
  Chanfray, and Marteau}}]{Martini:2009uj}
\bibinfo{author}{\bibfnamefont{M.}~\bibnamefont{Martini}},
  \bibinfo{author}{\bibfnamefont{M.}~\bibnamefont{Ericson}},
  \bibinfo{author}{\bibfnamefont{G.}~\bibnamefont{Chanfray}}, \bibnamefont{and}
  \bibinfo{author}{\bibfnamefont{J.}~\bibnamefont{Marteau}},
  \bibinfo{journal}{Phys.Rev.} \textbf{\bibinfo{volume}{C80}},
  \bibinfo{pages}{065501} (\bibinfo{year}{2009}), \eprint{0910.2622}.

\bibitem[{\citenamefont{Moniz et~al.}(1971)\citenamefont{Moniz, Sick, Whitney,
  Ficenec, Kephart et~al.}}]{Moniz:1971mt}
\bibinfo{author}{\bibfnamefont{E.}~\bibnamefont{Moniz}},
  \bibinfo{author}{\bibfnamefont{I.}~\bibnamefont{Sick}},
  \bibinfo{author}{\bibfnamefont{R.}~\bibnamefont{Whitney}},
  \bibinfo{author}{\bibfnamefont{J.}~\bibnamefont{Ficenec}},
  \bibinfo{author}{\bibfnamefont{R.~D.} \bibnamefont{Kephart}},
  \bibnamefont{et~al.}, \bibinfo{journal}{Phys.Rev.Lett.}
  \textbf{\bibinfo{volume}{26}}, \bibinfo{pages}{445} (\bibinfo{year}{1971}).

\bibitem[{\citenamefont{Van~Orden and Donnelly}(1981)}]{VanOrden:1980tg}
\bibinfo{author}{\bibfnamefont{J.}~\bibnamefont{Van~Orden}} \bibnamefont{and}
  \bibinfo{author}{\bibfnamefont{T.}~\bibnamefont{Donnelly}},
  \bibinfo{journal}{Annals Phys.} \textbf{\bibinfo{volume}{131}},
  \bibinfo{pages}{451} (\bibinfo{year}{1981}).

\bibitem[{\citenamefont{Leitner and Mosel}(2010)}]{Leitner:2010kp}
\bibinfo{author}{\bibfnamefont{T.}~\bibnamefont{Leitner}} \bibnamefont{and}
  \bibinfo{author}{\bibfnamefont{U.}~\bibnamefont{Mosel}},
  \bibinfo{journal}{Phys.Rev.} \textbf{\bibinfo{volume}{C81}},
  \bibinfo{pages}{064614} (\bibinfo{year}{2010}), \eprint{1004.4433}.

\bibitem[{\citenamefont{Drakoulakos et~al.}(2004)}]{Drakoulakos:2004gn}
\bibinfo{author}{\bibfnamefont{D.}~\bibnamefont{Drakoulakos}}
  \bibnamefont{et~al.} (\bibinfo{collaboration}{Minerva Collaboration})
  (\bibinfo{year}{2004}), \eprint{hep-ex/0405002}.

\bibitem[{\citenamefont{Adamson et~al.}(2008)}]{Adamson:2007gu}
\bibinfo{author}{\bibfnamefont{P.}~\bibnamefont{Adamson}} \bibnamefont{et~al.}
  (\bibinfo{collaboration}{MINOS Collaboration}), \bibinfo{journal}{Phys.Rev.}
  \textbf{\bibinfo{volume}{D77}}, \bibinfo{pages}{072002}
  (\bibinfo{year}{2008}), \eprint{0711.0769}.

\bibitem[{\citenamefont{Chen et~al.}(2007)}]{Chen:2007ae}
\bibinfo{author}{\bibfnamefont{H.}~\bibnamefont{Chen}} \bibnamefont{et~al.}
  (\bibinfo{collaboration}{MicroBooNE Collaboration}) (\bibinfo{year}{2007}),
  \bibinfo{note}{{FERMILAB-PROPOSAL-0974}}.

\bibitem[{\citenamefont{Ayres et~al.}(2004)}]{Ayres:2004js}
\bibinfo{author}{\bibfnamefont{D.}~\bibnamefont{Ayres}} \bibnamefont{et~al.}
  (\bibinfo{collaboration}{NOvA Collaboration}) (\bibinfo{year}{2004}),
  \eprint{hep-ex/0503053}.

\bibitem[{CDR()}]{CDR}
\bibinfo{note}{LBNE Conceptual Design Report from Oct 2012, volume 1},
  \urlprefix\url{http://lbne.fnal.gov/papers.shtml}.

\bibitem[{\citenamefont{Adams et~al.}(2013)}]{Adams:2013qkq}
\bibinfo{author}{\bibfnamefont{C.}~\bibnamefont{Adams}} \bibnamefont{et~al.}
  (\bibinfo{collaboration}{LBNE Collaboration}) (\bibinfo{year}{2013}),
  \eprint{1307.7335}.

\bibitem[{\citenamefont{Hayato}(2009)}]{Hayato:2009zz}
\bibinfo{author}{\bibfnamefont{Y.}~\bibnamefont{Hayato}},
  \bibinfo{journal}{Acta Phys.Polon.} \textbf{\bibinfo{volume}{B40}},
  \bibinfo{pages}{2477} (\bibinfo{year}{2009}).

\bibitem[{\citenamefont{Smith and Moniz}(1972)}]{Smith:1972xh}
\bibinfo{author}{\bibfnamefont{R.}~\bibnamefont{Smith}} \bibnamefont{and}
  \bibinfo{author}{\bibfnamefont{E.}~\bibnamefont{Moniz}},
  \bibinfo{journal}{Nucl.Phys.} \textbf{\bibinfo{volume}{B43}},
  \bibinfo{pages}{605} (\bibinfo{year}{1972}).

\bibitem[{\citenamefont{Benhar et~al.}(2008)\citenamefont{Benhar, Day, and
  Sick}}]{Benhar:2006wy}
\bibinfo{author}{\bibfnamefont{O.}~\bibnamefont{Benhar}},
  \bibinfo{author}{\bibfnamefont{D.}~\bibnamefont{Day}}, \bibnamefont{and}
  \bibinfo{author}{\bibfnamefont{I.}~\bibnamefont{Sick}},
  \bibinfo{journal}{Rev.Mod.Phys.} \textbf{\bibinfo{volume}{80}},
  \bibinfo{pages}{189} (\bibinfo{year}{2008}), \eprint{nucl-ex/0603029}.

\bibitem[{\citenamefont{Benhar et~al.}(2005)\citenamefont{Benhar, Farina,
  Nakamura, Sakuda, and Seki}}]{Benhar:2005dj}
\bibinfo{author}{\bibfnamefont{O.}~\bibnamefont{Benhar}},
  \bibinfo{author}{\bibfnamefont{N.}~\bibnamefont{Farina}},
  \bibinfo{author}{\bibfnamefont{H.}~\bibnamefont{Nakamura}},
  \bibinfo{author}{\bibfnamefont{M.}~\bibnamefont{Sakuda}}, \bibnamefont{and}
  \bibinfo{author}{\bibfnamefont{R.}~\bibnamefont{Seki}},
  \bibinfo{journal}{Phys.Rev.} \textbf{\bibinfo{volume}{D72}},
  \bibinfo{pages}{053005} (\bibinfo{year}{2005}), \eprint{hep-ph/0506116}.

\bibitem[{\citenamefont{Benhar and Meloni}(2007)}]{Benhar:2006nr}
\bibinfo{author}{\bibfnamefont{O.}~\bibnamefont{Benhar}} \bibnamefont{and}
  \bibinfo{author}{\bibfnamefont{D.}~\bibnamefont{Meloni}},
  \bibinfo{journal}{Nucl.Phys.} \textbf{\bibinfo{volume}{A789}},
  \bibinfo{pages}{379} (\bibinfo{year}{2007}), \eprint{hep-ph/0610403}.

\bibitem[{\citenamefont{Bodek and Ritchie}(1981)}]{Bodek:1981wr}
\bibinfo{author}{\bibfnamefont{A.}~\bibnamefont{Bodek}} \bibnamefont{and}
  \bibinfo{author}{\bibfnamefont{J.}~\bibnamefont{Ritchie}},
  \bibinfo{journal}{Phys.Rev.} \textbf{\bibinfo{volume}{D24}},
  \bibinfo{pages}{1400} (\bibinfo{year}{1981}).

\bibitem[{\citenamefont{Leitner
  et~al.}(2009{\natexlab{b}})\citenamefont{Leitner, Buss, Alvarez-Ruso, and
  Mosel}}]{Leitner:2008ue}
\bibinfo{author}{\bibfnamefont{T.}~\bibnamefont{Leitner}},
  \bibinfo{author}{\bibfnamefont{O.}~\bibnamefont{Buss}},
  \bibinfo{author}{\bibfnamefont{L.}~\bibnamefont{Alvarez-Ruso}},
  \bibnamefont{and} \bibinfo{author}{\bibfnamefont{U.}~\bibnamefont{Mosel}},
  \bibinfo{journal}{Phys.Rev.} \textbf{\bibinfo{volume}{C79}},
  \bibinfo{pages}{034601} (\bibinfo{year}{2009}{\natexlab{b}}),
  \eprint{0812.0587}.

\bibitem[{\citenamefont{Bradford et~al.}(2006)\citenamefont{Bradford, Bodek,
  Budd, and Arrington}}]{Bradford:2006yz}
\bibinfo{author}{\bibfnamefont{R.}~\bibnamefont{Bradford}},
  \bibinfo{author}{\bibfnamefont{A.}~\bibnamefont{Bodek}},
  \bibinfo{author}{\bibfnamefont{H.~S.} \bibnamefont{Budd}}, \bibnamefont{and}
  \bibinfo{author}{\bibfnamefont{J.}~\bibnamefont{Arrington}},
  \bibinfo{journal}{Nucl.Phys.Proc.Suppl.} \textbf{\bibinfo{volume}{159}},
  \bibinfo{pages}{127} (\bibinfo{year}{2006}), \eprint{hep-ex/0602017}.

\bibitem[{\citenamefont{Bodek et~al.}(2008)\citenamefont{Bodek, Avvakumov,
  Bradford, and Budd}}]{Bodek:2007ym}
\bibinfo{author}{\bibfnamefont{A.}~\bibnamefont{Bodek}},
  \bibinfo{author}{\bibfnamefont{S.}~\bibnamefont{Avvakumov}},
  \bibinfo{author}{\bibfnamefont{R.}~\bibnamefont{Bradford}}, \bibnamefont{and}
  \bibinfo{author}{\bibfnamefont{H.~S.} \bibnamefont{Budd}},
  \bibinfo{journal}{Eur.Phys.J.} \textbf{\bibinfo{volume}{C53}},
  \bibinfo{pages}{349} (\bibinfo{year}{2008}), \eprint{0708.1946}.

\bibitem[{\citenamefont{Leitner et~al.}(2006)\citenamefont{Leitner,
  Alvarez-Ruso, and Mosel}}]{Leitner:2006ww}
\bibinfo{author}{\bibfnamefont{T.}~\bibnamefont{Leitner}},
  \bibinfo{author}{\bibfnamefont{L.}~\bibnamefont{Alvarez-Ruso}},
  \bibnamefont{and} \bibinfo{author}{\bibfnamefont{U.}~\bibnamefont{Mosel}},
  \bibinfo{journal}{Phys.Rev.} \textbf{\bibinfo{volume}{C73}},
  \bibinfo{pages}{065502} (\bibinfo{year}{2006}), \eprint{nucl-th/0601103}.

\bibitem[{\citenamefont{Drechsel and Tiator}(1992)}]{Drechsel:1992pn}
\bibinfo{author}{\bibfnamefont{D.}~\bibnamefont{Drechsel}} \bibnamefont{and}
  \bibinfo{author}{\bibfnamefont{L.}~\bibnamefont{Tiator}},
  \bibinfo{journal}{J.Phys.} \textbf{\bibinfo{volume}{G18}},
  \bibinfo{pages}{449} (\bibinfo{year}{1992}).

\bibitem[{\citenamefont{Drechsel et~al.}(1999)\citenamefont{Drechsel, Hanstein,
  Kamalov, and Tiator}}]{Drechsel:1998hk}
\bibinfo{author}{\bibfnamefont{D.}~\bibnamefont{Drechsel}},
  \bibinfo{author}{\bibfnamefont{O.}~\bibnamefont{Hanstein}},
  \bibinfo{author}{\bibfnamefont{S.}~\bibnamefont{Kamalov}}, \bibnamefont{and}
  \bibinfo{author}{\bibfnamefont{L.}~\bibnamefont{Tiator}},
  \bibinfo{journal}{Nucl.Phys.} \textbf{\bibinfo{volume}{A645}},
  \bibinfo{pages}{145} (\bibinfo{year}{1999}), \eprint{nucl-th/9807001}.

\bibitem[{\citenamefont{Alvarez-Ruso et~al.}(1999)\citenamefont{Alvarez-Ruso,
  Singh, and Vicente~Vacas}}]{AlvarezRuso:1998hi}
\bibinfo{author}{\bibfnamefont{L.}~\bibnamefont{Alvarez-Ruso}},
  \bibinfo{author}{\bibfnamefont{S.}~\bibnamefont{Singh}}, \bibnamefont{and}
  \bibinfo{author}{\bibfnamefont{M.}~\bibnamefont{Vicente~Vacas}},
  \bibinfo{journal}{Phys.Rev.} \textbf{\bibinfo{volume}{C59}},
  \bibinfo{pages}{3386} (\bibinfo{year}{1999}), \eprint{nucl-th/9804007}.

\bibitem[{\citenamefont{Rein and Sehgal}(1981)}]{Rein:1981ys}
\bibinfo{author}{\bibfnamefont{D.}~\bibnamefont{Rein}} \bibnamefont{and}
  \bibinfo{author}{\bibfnamefont{L.}~\bibnamefont{Sehgal}},
  \bibinfo{journal}{Phys.Lett.} \textbf{\bibinfo{volume}{B104}},
  \bibinfo{pages}{394} (\bibinfo{year}{1981}).

\bibitem[{\citenamefont{Dytman}(2009)}]{Dytman:2009zz}
\bibinfo{author}{\bibfnamefont{S.}~\bibnamefont{Dytman}},
  \bibinfo{journal}{Acta Phys.Polon.} \textbf{\bibinfo{volume}{B40}},
  \bibinfo{pages}{2445} (\bibinfo{year}{2009}).

\bibitem[{\citenamefont{Martini et~al.}(2010)\citenamefont{Martini, Ericson,
  Chanfray, and Marteau}}]{Martini:2010ex}
\bibinfo{author}{\bibfnamefont{M.}~\bibnamefont{Martini}},
  \bibinfo{author}{\bibfnamefont{M.}~\bibnamefont{Ericson}},
  \bibinfo{author}{\bibfnamefont{G.}~\bibnamefont{Chanfray}}, \bibnamefont{and}
  \bibinfo{author}{\bibfnamefont{J.}~\bibnamefont{Marteau}},
  \bibinfo{journal}{Phys.Rev.} \textbf{\bibinfo{volume}{C81}},
  \bibinfo{pages}{045502} (\bibinfo{year}{2010}), \eprint{1002.4538}.

\bibitem[{\citenamefont{Benhar et~al.}(2010)\citenamefont{Benhar, Coletti, and
  Meloni}}]{Benhar:2010nx}
\bibinfo{author}{\bibfnamefont{O.}~\bibnamefont{Benhar}},
  \bibinfo{author}{\bibfnamefont{P.}~\bibnamefont{Coletti}}, \bibnamefont{and}
  \bibinfo{author}{\bibfnamefont{D.}~\bibnamefont{Meloni}},
  \bibinfo{journal}{Phys.Rev.Lett.} \textbf{\bibinfo{volume}{105}},
  \bibinfo{pages}{132301} (\bibinfo{year}{2010}), \eprint{1006.4783}.

\bibitem[{\citenamefont{Amaro et~al.}(2011)\citenamefont{Amaro, Barbaro,
  Caballero, Donnelly, and Williamson}}]{Amaro:2010sd}
\bibinfo{author}{\bibfnamefont{J.}~\bibnamefont{Amaro}},
  \bibinfo{author}{\bibfnamefont{M.}~\bibnamefont{Barbaro}},
  \bibinfo{author}{\bibfnamefont{J.}~\bibnamefont{Caballero}},
  \bibinfo{author}{\bibfnamefont{T.}~\bibnamefont{Donnelly}}, \bibnamefont{and}
  \bibinfo{author}{\bibfnamefont{C.}~\bibnamefont{Williamson}},
  \bibinfo{journal}{Phys.Lett.} \textbf{\bibinfo{volume}{B696}},
  \bibinfo{pages}{151} (\bibinfo{year}{2011}), \eprint{1010.1708}.

\bibitem[{\citenamefont{Juszczak et~al.}(2010)\citenamefont{Juszczak, Sobczyk,
  and Zmuda}}]{Juszczak:2010ve}
\bibinfo{author}{\bibfnamefont{C.}~\bibnamefont{Juszczak}},
  \bibinfo{author}{\bibfnamefont{J.~T.} \bibnamefont{Sobczyk}},
  \bibnamefont{and} \bibinfo{author}{\bibfnamefont{J.}~\bibnamefont{Zmuda}},
  \bibinfo{journal}{Phys.Rev.} \textbf{\bibinfo{volume}{C82}},
  \bibinfo{pages}{045502} (\bibinfo{year}{2010}), \eprint{1007.2195}.

\bibitem[{\citenamefont{Bodek et~al.}(2011)\citenamefont{Bodek, Budd, and
  Christy}}]{Bodek:2011ps}
\bibinfo{author}{\bibfnamefont{A.}~\bibnamefont{Bodek}},
  \bibinfo{author}{\bibfnamefont{H.}~\bibnamefont{Budd}}, \bibnamefont{and}
  \bibinfo{author}{\bibfnamefont{M.}~\bibnamefont{Christy}},
  \bibinfo{journal}{Eur.Phys.J.} \textbf{\bibinfo{volume}{C71}},
  \bibinfo{pages}{1726} (\bibinfo{year}{2011}), \eprint{1106.0340}.

\bibitem[{\citenamefont{Nieves et~al.}(2011)\citenamefont{Nieves, Ruiz~Simo,
  and Vicente~Vacas}}]{Nieves:2011pp}
\bibinfo{author}{\bibfnamefont{J.}~\bibnamefont{Nieves}},
  \bibinfo{author}{\bibfnamefont{I.}~\bibnamefont{Ruiz~Simo}},
  \bibnamefont{and}
  \bibinfo{author}{\bibfnamefont{M.}~\bibnamefont{Vicente~Vacas}},
  \bibinfo{journal}{Phys.Rev.} \textbf{\bibinfo{volume}{C83}},
  \bibinfo{pages}{045501} (\bibinfo{year}{2011}), \eprint{1102.2777}.

\bibitem[{\citenamefont{Nieves et~al.}(2012{\natexlab{b}})\citenamefont{Nieves,
  Ruiz~Simo, and Vicente~Vacas}}]{Nieves:2011yp}
\bibinfo{author}{\bibfnamefont{J.}~\bibnamefont{Nieves}},
  \bibinfo{author}{\bibfnamefont{I.}~\bibnamefont{Ruiz~Simo}},
  \bibnamefont{and}
  \bibinfo{author}{\bibfnamefont{M.}~\bibnamefont{Vicente~Vacas}},
  \bibinfo{journal}{Phys.Lett.} \textbf{\bibinfo{volume}{B707}},
  \bibinfo{pages}{72} (\bibinfo{year}{2012}{\natexlab{b}}), \eprint{1106.5374}.

\bibitem[{\citenamefont{Martini et~al.}(2011)\citenamefont{Martini, Ericson,
  and Chanfray}}]{Martini:2011wp}
\bibinfo{author}{\bibfnamefont{M.}~\bibnamefont{Martini}},
  \bibinfo{author}{\bibfnamefont{M.}~\bibnamefont{Ericson}}, \bibnamefont{and}
  \bibinfo{author}{\bibfnamefont{G.}~\bibnamefont{Chanfray}},
  \bibinfo{journal}{Phys.Rev.} \textbf{\bibinfo{volume}{C84}},
  \bibinfo{pages}{055502} (\bibinfo{year}{2011}), \eprint{1110.0221}.

\bibitem[{\citenamefont{Lalakulich
  et~al.}(2012{\natexlab{b}})\citenamefont{Lalakulich, Gallmeister, and
  Mosel}}]{Lalakulich:2012ac}
\bibinfo{author}{\bibfnamefont{O.}~\bibnamefont{Lalakulich}},
  \bibinfo{author}{\bibfnamefont{K.}~\bibnamefont{Gallmeister}},
  \bibnamefont{and} \bibinfo{author}{\bibfnamefont{U.}~\bibnamefont{Mosel}},
  \bibinfo{journal}{Phys.Rev.} \textbf{\bibinfo{volume}{C86}},
  \bibinfo{pages}{014614} (\bibinfo{year}{2012}{\natexlab{b}}),
  \eprint{1203.2935}.

\bibitem[{\citenamefont{Benhar et~al.}(2013)\citenamefont{Benhar, Lovato, and
  Rocco}}]{Benhar:2013bba}
\bibinfo{author}{\bibfnamefont{O.}~\bibnamefont{Benhar}},
  \bibinfo{author}{\bibfnamefont{A.}~\bibnamefont{Lovato}}, \bibnamefont{and}
  \bibinfo{author}{\bibfnamefont{N.}~\bibnamefont{Rocco}}
  (\bibinfo{year}{2013}), \eprint{1312.1210}.

\bibitem[{\citenamefont{Benhar and Rocco}(2013)}]{Benhar:2013bwa}
\bibinfo{author}{\bibfnamefont{O.}~\bibnamefont{Benhar}} \bibnamefont{and}
  \bibinfo{author}{\bibfnamefont{N.}~\bibnamefont{Rocco}}
  (\bibinfo{year}{2013}), \eprint{1310.3869}.

\bibitem[{\citenamefont{Gran et~al.}(2013)\citenamefont{Gran, Nieves, Sanchez,
  and Vicente~Vacas}}]{Gran:2013kda}
\bibinfo{author}{\bibfnamefont{R.}~\bibnamefont{Gran}},
  \bibinfo{author}{\bibfnamefont{J.}~\bibnamefont{Nieves}},
  \bibinfo{author}{\bibfnamefont{F.}~\bibnamefont{Sanchez}}, \bibnamefont{and}
  \bibinfo{author}{\bibfnamefont{M.}~\bibnamefont{Vicente~Vacas}},
  \bibinfo{journal}{Phys.Rev.} \textbf{\bibinfo{volume}{D88}},
  \bibinfo{pages}{113007} (\bibinfo{year}{2013}), \eprint{1307.8105}.

\bibitem[{\citenamefont{Aguilar-Arevalo et~al.}(2010)}]{AguilarArevalo:2010zc}
\bibinfo{author}{\bibfnamefont{A.}~\bibnamefont{Aguilar-Arevalo}}
  \bibnamefont{et~al.} (\bibinfo{collaboration}{MiniBooNE Collaboration}),
  \bibinfo{journal}{Phys.Rev.} \textbf{\bibinfo{volume}{D81}},
  \bibinfo{pages}{092005} (\bibinfo{year}{2010}), \eprint{1002.2680}.

\bibitem[{\citenamefont{Katori}(2013)}]{Katori:2013eoa}
\bibinfo{author}{\bibfnamefont{T.}~\bibnamefont{Katori}}
  (\bibinfo{year}{2013}), \eprint{1304.6014}.

\bibitem[{\citenamefont{Boyd et~al.}(2009)\citenamefont{Boyd, Dytman,
  Hernandez, Sobczyk, and Tacik}}]{Boyd:2009zz}
\bibinfo{author}{\bibfnamefont{S.}~\bibnamefont{Boyd}},
  \bibinfo{author}{\bibfnamefont{S.}~\bibnamefont{Dytman}},
  \bibinfo{author}{\bibfnamefont{E.}~\bibnamefont{Hernandez}},
  \bibinfo{author}{\bibfnamefont{J.}~\bibnamefont{Sobczyk}}, \bibnamefont{and}
  \bibinfo{author}{\bibfnamefont{R.}~\bibnamefont{Tacik}},
  \bibinfo{journal}{AIP Conf.Proc.} \textbf{\bibinfo{volume}{1189}},
  \bibinfo{pages}{60} (\bibinfo{year}{2009}).

\bibitem[{\citenamefont{Nakamura et~al.}(2010)\citenamefont{Nakamura, Sato,
  Lee, Szczerbinska, and Kubodera}}]{Nakamura:2009iq}
\bibinfo{author}{\bibfnamefont{S.}~\bibnamefont{Nakamura}},
  \bibinfo{author}{\bibfnamefont{T.}~\bibnamefont{Sato}},
  \bibinfo{author}{\bibfnamefont{T.-S.} \bibnamefont{Lee}},
  \bibinfo{author}{\bibfnamefont{B.}~\bibnamefont{Szczerbinska}},
  \bibnamefont{and} \bibinfo{author}{\bibfnamefont{K.}~\bibnamefont{Kubodera}},
  \bibinfo{journal}{Phys.Rev.} \textbf{\bibinfo{volume}{C81}},
  \bibinfo{pages}{035502} (\bibinfo{year}{2010}), \eprint{0910.1057}.

\bibitem[{\citenamefont{Nakamura}(2013)}]{Nakamura:2011rt}
\bibinfo{author}{\bibfnamefont{S.~X.} \bibnamefont{Nakamura}},
  \bibinfo{journal}{J.Phys.Conf.Ser.} \textbf{\bibinfo{volume}{408}},
  \bibinfo{pages}{012043} (\bibinfo{year}{2013}), \eprint{1109.4443}.

\bibitem[{\citenamefont{Rein and Sehgal}(1983)}]{Rein:1982pf}
\bibinfo{author}{\bibfnamefont{D.}~\bibnamefont{Rein}} \bibnamefont{and}
  \bibinfo{author}{\bibfnamefont{L.~M.} \bibnamefont{Sehgal}},
  \bibinfo{journal}{Nucl.Phys.} \textbf{\bibinfo{volume}{B223}},
  \bibinfo{pages}{29} (\bibinfo{year}{1983}).

\bibitem[{\citenamefont{Leitner
  et~al.}(2009{\natexlab{c}})\citenamefont{Leitner, Mosel, and
  Winkelmann}}]{Leitner:2009ph}
\bibinfo{author}{\bibfnamefont{T.}~\bibnamefont{Leitner}},
  \bibinfo{author}{\bibfnamefont{U.}~\bibnamefont{Mosel}}, \bibnamefont{and}
  \bibinfo{author}{\bibfnamefont{S.}~\bibnamefont{Winkelmann}},
  \bibinfo{journal}{Phys.Rev.} \textbf{\bibinfo{volume}{C79}},
  \bibinfo{pages}{057601} (\bibinfo{year}{2009}{\natexlab{c}}),
  \eprint{0901.2837}.

\bibitem[{\citenamefont{Huber et~al.}(2009)\citenamefont{Huber, Lindner,
  Schwetz, and Winter}}]{Huber:2009cw}
\bibinfo{author}{\bibfnamefont{P.}~\bibnamefont{Huber}},
  \bibinfo{author}{\bibfnamefont{M.}~\bibnamefont{Lindner}},
  \bibinfo{author}{\bibfnamefont{T.}~\bibnamefont{Schwetz}}, \bibnamefont{and}
  \bibinfo{author}{\bibfnamefont{W.}~\bibnamefont{Winter}},
  \bibinfo{journal}{JHEP} \textbf{\bibinfo{volume}{0911}}, \bibinfo{pages}{044}
  (\bibinfo{year}{2009}), \eprint{0907.1896}.

\bibitem[{\citenamefont{Bernard et~al.}(2002)\citenamefont{Bernard,
  Elouadrhiri, and Meissner}}]{Bernard:2001rs}
\bibinfo{author}{\bibfnamefont{V.}~\bibnamefont{Bernard}},
  \bibinfo{author}{\bibfnamefont{L.}~\bibnamefont{Elouadrhiri}},
  \bibnamefont{and} \bibinfo{author}{\bibfnamefont{U.}~\bibnamefont{Meissner}},
  \bibinfo{journal}{J.Phys.} \textbf{\bibinfo{volume}{G28}},
  \bibinfo{pages}{R1} (\bibinfo{year}{2002}), \eprint{hep-ph/0107088}.

\bibitem[{\citenamefont{Gonzalez-Garcia
  et~al.}(2012)\citenamefont{Gonzalez-Garcia, Maltoni, Salvado, and
  Schwetz}}]{GonzalezGarcia:2012sz}
\bibinfo{author}{\bibfnamefont{M.}~\bibnamefont{Gonzalez-Garcia}},
  \bibinfo{author}{\bibfnamefont{M.}~\bibnamefont{Maltoni}},
  \bibinfo{author}{\bibfnamefont{J.}~\bibnamefont{Salvado}}, \bibnamefont{and}
  \bibinfo{author}{\bibfnamefont{T.}~\bibnamefont{Schwetz}},
  \bibinfo{journal}{JHEP} \textbf{\bibinfo{volume}{1212}}, \bibinfo{pages}{123}
  (\bibinfo{year}{2012}), \eprint{1209.3023}.

\bibitem[{\citenamefont{Fiorentini et~al.}(2013)}]{Fiorentini:2013ezn}
\bibinfo{author}{\bibfnamefont{G.}~\bibnamefont{Fiorentini}}
  \bibnamefont{et~al.} (\bibinfo{collaboration}{MINERvA Collaboration}),
  \bibinfo{journal}{Phys.Rev.Lett.} \textbf{\bibinfo{volume}{111}},
  \bibinfo{pages}{022502} (\bibinfo{year}{2013}), \eprint{1305.2243}.

\bibitem[{\citenamefont{Nieves et~al.}(2013)\citenamefont{Nieves, Ruiz~Simo,
  and Vicente~Vacas}}]{Nieves:2013fr}
\bibinfo{author}{\bibfnamefont{J.}~\bibnamefont{Nieves}},
  \bibinfo{author}{\bibfnamefont{I.}~\bibnamefont{Ruiz~Simo}},
  \bibnamefont{and}
  \bibinfo{author}{\bibfnamefont{M.}~\bibnamefont{Vicente~Vacas}},
  \bibinfo{journal}{Phys.Lett.} \textbf{\bibinfo{volume}{B721}},
  \bibinfo{pages}{90} (\bibinfo{year}{2013}), \eprint{1302.0703}.

\bibitem[{\citenamefont{Aguilar-Arevalo et~al.}(2013)}]{AguilarArevalo:2013hm}
\bibinfo{author}{\bibfnamefont{A.}~\bibnamefont{Aguilar-Arevalo}}
  \bibnamefont{et~al.} (\bibinfo{collaboration}{MiniBooNE Collaboration}),
  \bibinfo{journal}{Phys.Rev.} \textbf{\bibinfo{volume}{D88}},
  \bibinfo{pages}{032001} (\bibinfo{year}{2013}), \eprint{1301.7067}.

\bibitem[{\citenamefont{Fields et~al.}(2013)}]{Fields:2013zhk}
\bibinfo{author}{\bibfnamefont{L.}~\bibnamefont{Fields}} \bibnamefont{et~al.}
  (\bibinfo{collaboration}{MINERvA Collaboration}),
  \bibinfo{journal}{Phys.Rev.Lett.} \textbf{\bibinfo{volume}{111}},
  \bibinfo{pages}{022501} (\bibinfo{year}{2013}), \eprint{1305.2234}.

\bibitem[{\citenamefont{Abe et~al.}(2011{\natexlab{b}})\citenamefont{Abe, Abe,
  Aihara, Fukuda, Hayato et~al.}}]{Abe:2011ts}
\bibinfo{author}{\bibfnamefont{K.}~\bibnamefont{Abe}},
  \bibinfo{author}{\bibfnamefont{T.}~\bibnamefont{Abe}},
  \bibinfo{author}{\bibfnamefont{H.}~\bibnamefont{Aihara}},
  \bibinfo{author}{\bibfnamefont{Y.}~\bibnamefont{Fukuda}},
  \bibinfo{author}{\bibfnamefont{Y.}~\bibnamefont{Hayato}},
  \bibnamefont{et~al.} (\bibinfo{year}{2011}{\natexlab{b}}),
  \eprint{1109.3262}.

\bibitem[{\citenamefont{Baussan et~al.}(2013)}]{Baussan:2013zcy}
\bibinfo{author}{\bibfnamefont{E.}~\bibnamefont{Baussan}} \bibnamefont{et~al.}
  (\bibinfo{collaboration}{ESSnuSB Collaboration}) (\bibinfo{year}{2013}),
  \eprint{1309.7022}.

\bibitem[{\citenamefont{Stahl et~al.}(2012)\citenamefont{Stahl, Wiebusch,
  Guler, Kamiscioglu, Sever et~al.}}]{Stahl:2012exa}
\bibinfo{author}{\bibfnamefont{A.}~\bibnamefont{Stahl}},
  \bibinfo{author}{\bibfnamefont{C.}~\bibnamefont{Wiebusch}},
  \bibinfo{author}{\bibfnamefont{A.}~\bibnamefont{Guler}},
  \bibinfo{author}{\bibfnamefont{M.}~\bibnamefont{Kamiscioglu}},
  \bibinfo{author}{\bibfnamefont{R.}~\bibnamefont{Sever}}, \bibnamefont{et~al.}
  (\bibinfo{year}{2012}), \bibinfo{note}{{CERN-SPSC-2012-021, SPSC-EOI-007}}.

\bibitem[{\citenamefont{Amaro et~al.}(2012)\citenamefont{Amaro, Barbaro,
  Caballero, and Donnelly}}]{Amaro:2011aa}
\bibinfo{author}{\bibfnamefont{J.}~\bibnamefont{Amaro}},
  \bibinfo{author}{\bibfnamefont{M.}~\bibnamefont{Barbaro}},
  \bibinfo{author}{\bibfnamefont{J.}~\bibnamefont{Caballero}},
  \bibnamefont{and} \bibinfo{author}{\bibfnamefont{T.}~\bibnamefont{Donnelly}},
  \bibinfo{journal}{Phys.Rev.Lett.} \textbf{\bibinfo{volume}{108}},
  \bibinfo{pages}{152501} (\bibinfo{year}{2012}), \eprint{1112.2123}.

\bibitem[{\citenamefont{Martini and Ericson}(2013)}]{Martini:2013sha}
\bibinfo{author}{\bibfnamefont{M.}~\bibnamefont{Martini}} \bibnamefont{and}
  \bibinfo{author}{\bibfnamefont{M.}~\bibnamefont{Ericson}},
  \bibinfo{journal}{Phys.Rev.} \textbf{\bibinfo{volume}{C87}},
  \bibinfo{pages}{065501} (\bibinfo{year}{2013}), \eprint{1303.7199}.

\bibitem[{\citenamefont{Abe et~al.}(2013)}]{Abe:2013xua}
\bibinfo{author}{\bibfnamefont{K.}~\bibnamefont{Abe}} \bibnamefont{et~al.}
  (\bibinfo{collaboration}{T2K Collaboration}), \bibinfo{journal}{Phys.Rev.}
  \textbf{\bibinfo{volume}{D88}}, \bibinfo{pages}{032002}
  (\bibinfo{year}{2013}), \eprint{1304.0841}.

\bibitem[{\citenamefont{Huber et~al.}(2005)\citenamefont{Huber, Lindner, and
  Winter}}]{Huber:2004ka}
\bibinfo{author}{\bibfnamefont{P.}~\bibnamefont{Huber}},
  \bibinfo{author}{\bibfnamefont{M.}~\bibnamefont{Lindner}}, \bibnamefont{and}
  \bibinfo{author}{\bibfnamefont{W.}~\bibnamefont{Winter}},
  \bibinfo{journal}{Comput.Phys.Commun.} \textbf{\bibinfo{volume}{167}},
  \bibinfo{pages}{195} (\bibinfo{year}{2005}), \eprint{hep-ph/0407333}.

\bibitem[{\citenamefont{Huber et~al.}(2007)\citenamefont{Huber, Kopp, Lindner,
  Rolinec, and Winter}}]{Huber:2007ji}
\bibinfo{author}{\bibfnamefont{P.}~\bibnamefont{Huber}},
  \bibinfo{author}{\bibfnamefont{J.}~\bibnamefont{Kopp}},
  \bibinfo{author}{\bibfnamefont{M.}~\bibnamefont{Lindner}},
  \bibinfo{author}{\bibfnamefont{M.}~\bibnamefont{Rolinec}}, \bibnamefont{and}
  \bibinfo{author}{\bibfnamefont{W.}~\bibnamefont{Winter}},
  \bibinfo{journal}{Comput.Phys.Commun.} \textbf{\bibinfo{volume}{177}},
  \bibinfo{pages}{432} (\bibinfo{year}{2007}), \eprint{hep-ph/0701187}.

\end{thebibliography}

\end{document}